\RequirePackage{rotating}
\documentclass[useAMS,usenatbib]{mn2e}
\usepackage{graphicx}
\usepackage{rotating}
\usepackage{amssymb}
\usepackage{savesym}
\usepackage{amsmath}
\savesymbol{iint}
\usepackage{txfonts}
\restoresymbol{TXF}{iint}\usepackage{mathtools}

\title[Optical and near-infrared observations of SN 2014ck]{{Optical and near infrared observations of SN 2014ck: an outlier among the Type Iax supernovae}
}
\author[L. Tomasella et al.]
  {L.~Tomasella,$^1$
E.~Cappellaro,$^1$ S.~Benetti,$^1$ A.~Pastorello,$^1$ E.Y.~Hsiao,$^{2,3}$ D.J.~Sand,$^4$ \newauthor M.~Stritzinger,$^2$ S.~Valenti,$^{5,6}$ C.~McCully,$^{5,6}$ I.~Arcavi,$^{6,7}$
N.~Elias-Rosa,$^1$  J.~Harmanen,$^8$ \newauthor A.~Harutyunyan,$^9$ G.~Hosseinzadeh,$^{5,6}$ D.A.~Howell,$^{5,6}$ E.~Kankare,$^{10}$ 
 \newauthor A. Morales-Garoffolo,$^{11}$  F.~Taddia,$^{12}$ L. Tartaglia,$^1$ G. Terreran,$^{1,10}$ M.~Turatto$^1$  \\
  $^1$INAF, Osservatorio Astronomico di Padova, 35122 Padova, Italy\\
  $^2$Department of Physics and Astronomy, Aarhus University, Ny Munkegade 120, 8000 Aarhus C, Denmark\\
  $^3$Department of Physics, Florida State University, 77 Chieftan Way, Tallahassee, FL 32306, USA\\
  $^4$Texas Tech University, Physics Department, Box 41051, Lubbock, TX 79409-1051, USA\\
  $^5$Department of Physics, University of California, Santa Barbara, Broida Hall, Mail Code 9530, Santa Barbara, CA 93106-9530, USA\\
  $^6$Las Cumbres Observatory Global Telescope Network, 6740 Cortona Dr., Suite 102, Goleta, CA 93117, USA\\ 
  $^7$Kavli Institute for Theoretical Physics, University of California, Santa Barbara, CA 93106, USA\\
  $^8$Tuorla Observatory, Department of Physics and Astronomy, University of Turku, V\"ais\"al\"antie 20, FI-21500, Piikki\"o, Finland\\  
  $^9$Fundaci\'on Galileo Galilei, INAF Telescopio Nazionale Galileo, Rambla Jos\'e Ana Fern\'andez P\'erez 7, 38712 Bre\~na Baja, TF, Spain\\
  $^{10}$Astrophysics Research Centre, School of Mathematics and Physics, Queen's University Belfast, BT7 1NN, UK\\
  $^{11}$Institut de Ci\`encies de l'Espai (CSIC-IEEC), Campus UAB, Cam\'i de Can Magrans S/N, 08193 Cerdanyola (Barcelona), Spain\\
  $^{12}$The Oskar Klein Centre, Department of Astronomy, AlbaNova, SE-106 91 Stockholm, Sweden\\  
    }
\date{Accepted 2016 March 22. Received 2016 March 9; in original form 2015 November 30}

\pagerange{\pageref{firstpage}--\pageref{lastpage}} \pubyear{2015}

\begin{document}

\label{firstpage}

\maketitle

\begin{abstract}

We present a comprehensive set of optical and near-infrared photometric and spectroscopic observations for SN~2014ck, extending from pre-maximum to six months later. These data indicate that SN~2014ck is photometrically nearly identical to SN~2002cx, which is the prototype of the class of peculiar transients named SNe~Iax. Similar to SN~2002cx, SN~2014ck reached a peak brightness $M_B=-17.37 \pm 0.15$~mag, with a post-maximum decline-rate 
$\Delta m_{15} (B) = 1.76 \pm 0.15$~mag. However, the spectroscopic sequence shows similarities with SN~2008ha, which was three magnitudes fainter and faster declining. In particular, SN~2014ck exhibits extremely low ejecta velocities, $\sim 3000$~km~s$^{-1}$ at maximum, which are close to the value measured for SN~2008ha and half the value inferred for SN~2002cx. 
The bolometric light curve of SN~2014ck is consistent with the production of $0.10^{+0.04}_{-0.03} M_{\odot}$ of $^{56}$Ni. 
The spectral identification of several iron-peak features, in particular Co~II lines in the NIR, provides a clear link to SNe~Ia. Also, the detection of narrow Si, S and C features in the pre-maximum spectra suggests a thermonuclear explosion mechanism. 
The late-phase spectra show a complex overlap of both permitted and forbidden Fe, Ca and Co lines. 
The appearance of strong [Ca~II]~$\lambda\lambda$7292, 7324 again mirrors the late-time spectra of SN~2008ha and SN~2002cx. 
The photometric resemblance to SN~2002cx and the spectral similarities to SN~2008ha highlight the peculiarity of SN~2014ck, and the complexity and heterogeneity of the SNe~Iax class.

\end{abstract}


\section{Introduction}

The discovery of several peculiar Type Ia supernovae (SNe~Ia) has drawn the attention both to the photometric and spectroscopic diversity among this class of otherwise homogeneous transients. The dispersion of the luminosity--decline rate relation \citep{phillips:1993,hamuy:1995,hamuya:1996} can be explained    
by an additional correlation between the decline rate and the colour at maximum light \citep{hamuyb:1996,tripp:1998,branch:1998,tripp:1999}. Hence SNe~Ia can be arranged into a photometric sequence extending from  luminous, blue, slowly declining SN~Ia, like 
SN~1991T, to normal events \citep{branch:1993}, and finally to sub-luminous, red, quickly declining objects, like SN~1991bg \citep{filippenkoa:1992,filippenkob:1992,lei:1993,turatto:1996}.  
SNe Ia also appear to form a spectroscopic sequence based on the
systematic variations in the flux ratios of several spectral features near maximum light \citep[e.g. Si~II $\lambda\lambda$5972, 6355, see][]{nugent:1995}.
The common view is, despite their diversity, peculiar events such as the luminous 1991T-like and sub-luminous 1991bg-like SNe~Ia, just like the normal population of SNe~Ia, 
originate from the thermonuclear explosion of a C/O white dwarf (WD) that exceeds the Chandrasekhar mass after accreting mass from a companion star in a binary system.

However, there is a group of peculiar SNe~Ia that challenges the canonical Chandrasekhar-mass explosion channel. The prototype of this class is SN~2002cx \citep{li:2003}, which shows a peak luminosity significantly lower than that of normal SNe~Ia, even though its light curve decline-rate parameter is comparable to normal events. Spectra obtained near maximum light resemble those of over-luminous 1991T-like objects (with a blue continuum and absorption from higher-ionisation species), even if a low ejecta velocity ($\sim6000$~km~s$^{-1}$ at the epoch of $B$-band maximum light) points towards a moderate kinetic energy from the explosion \citep{li:2003}. The late-time spectra show narrow iron and cobalt lines \citep{li:2003,jha:2006}, in stark contrast to normal SNe~Ia at similar epochs.  After the pioneering studies by \cite{li:2003} and \cite{jha:2006} on SN~2002cx, both new and old SN discoveries have been classified or reclassified as 2002cx-like events, and it has become clear these transients are not so rare. This class was labelled Type~Iax supernovae (SNe~Iax) by \cite{foley:2013}, who presented a review on the entire group and defined clear observational criteria to classify a Type~Iax event. 

A variety of explosion scenarios and potential progenitors or progenitor systems have been proposed to explain each event \citep[see][and references therein for a recent review]{liua:2015}.
Although the leading models for SNe~Iax are thermonuclear explosion of a C/O WD  \citep{foley:2009,jordan:2012,kromer:2013,fink:2014,stritz:2015,kromer:2015,liub:2015}, a core-collapse scenario has been proposed at least for SN~2008ha \citep{valenti:2009,foley:2009,moriya:2010}, which is the most extreme member of SN~Iax class to date. The latter however is controversial because of the detection of C/O burning products in the maximum-light spectrum of SN~2008ha \citep{foley:2009,foley:2010b}, providing a link to thermonuclear explosions.   
In principle, the best way to shed light on this issue would be the detection of a progenitor in pre-explosion images.  Recently, \cite{mccully:2014} reported the detection of a luminous blue source coincident (at the 0.8$\sigma$ level) with the location of Type~Iax SN~2012Z in {\em Hubble Space Telescope} ({\em HST}) pre-explosion images. 
Although the photometric properties of this object suggest a C/O WD primary plus a He-star companion progenitor system, the explosion of a single massive star has not definitely been ruled out.  In this case, post-explosion imaging, obtained after the SN fades away, should help to distinguish between the two models. 
For two other SNe~Iax, no sources were detected in pre-explosion images, but limits were obtained that exclude massive stars as potential progenitors \citep[SNe~2008ge and 2014dt, see][]{foley:2010a,foley:2015}.  

Given the diversity of this SN class, one may consider the possibility that multiple progenitor channels may lead to the production of SNe~Iax.  
In fact, the forty-some objects classified as SNe~Iax have a number of similarities, but also noteworthy differences. 
In particular, they show a large range in luminosity at maximum, from $M_{V}$ $\approx$ $-$14.2~mag of the faint SN~2008ha \citep{valenti:2009,foley:2009,stritz:2014} to $M_{V}$ $\approx$ $-$18.5~mag of SNe~2009ku \citep{narayan:2011} and 2012Z \citep{stritz:2015,yamanaka:2015}. The ejecta velocities near maximum brightness also exhibit a large spread, ranging from $\sim2000$ to $\sim8000$~km~s$^{-1}$. 
For the majority of SNe~Iax, there appears to be a correlation between ejecta velocity and peak luminosity, with the higher-velocity objects being also the brighter ones \citep{mcclelland:2010,foley:2013}. However, SN~2009ku, a low-velocity, high-luminosity SN studied by \cite{narayan:2011}, does not follow the trend \citep[note however that the first spectrum was taken long after maximum and so the inferred ejecta velocity is  uncertain, see][]{foley:2013}.  
In this paper, we present the results of a comprehensive observational campaign of SN~2014ck, which started well before maximum light. It turns out that SN~2014ck is an outlier among SNe~Iax, as it mirrors SN~2002cx from a photometric point of view, while the early spectra exhibit extremely narrow spectral lines, indicating very low expansion velocities of the ejecta.

This paper is organised as follows: in Section~2, we give some basic information about the SN discovery and the host galaxy, and we describe the follow-up campaign. In Section~3, we analyse {\it HST} pre-discovery images. 
We discuss data reduction and present the photometric evolution and visual and near-infrared spectroscopic sequences of SN~2014ck in Section~4. In Section~5 the Galactic and host galaxy reddening is estimated.  
Descriptions of the photometric and spectroscopic properties of SN~2014ck are reported in Sections~6 and 7, respectively. Expansion velocities of the ejecta, along with the photospheric temperatures, are deduced from the spectra. Spectral modelling with the {\tt SYNOW} code is used to assist in line identification. A final discussion of the available data in the context of the explosion models follows in Section~8.

\section{SN 2014ck discovery and follow-up observations}\label{discovery}

SN 2014ck was discovered by the Lick Observatory Supernova Search \citep[LOSS;][]{filippenko:2001}, on 2014 June 29.47 UT, at an apparent magnitude of 16.4~mag using the Katzman Automatic Imaging Telescope \citep[KAIT;][]{hayakawa:2014}. A marginal detection on 2014 June 24.5 UT was also reported by LOSS with an approximate $R$-band magnitude of 17.0~mag. However, a subsequent analysis of KAIT images on 2014 June 13, 23, 24, 25, 28 and 29, performed independently by the LOSS team (Zheng 2016, private communication) and by us\footnote{We thank WeiKang Zheng and Alex Filippenko for sending us LOSS/KAIT pre-discovery images.}, postpones this marginal detection by approximately one day (2014 June 25.5 UT, with an approximate $r$-band magnitude of 18.15$\pm$0.44 mag, as reported in Table~\ref{sloan}).

The SN is located 4\farcs3~E and 0\farcs5~S from the centre of the spiral galaxy UGC~12182 (Figure~1). 
An heliocentric recessional velocity of 1490~km~s$^{-1}$ for UGC~12182  is listed in the NASA/IPAC Extragalactic Database (NED), as taken from ``The Updated Zwicky Catalogue'' \citep{falco:1999}. The distance and distance modulus (adopting $H_0 = 73 \pm 5$~km~s$^{-1}$~Mpc$^{-1}$), corrected for the Virgo, Great Attractor and Shapley infall, are $24.4 \pm 1.7$~Mpc and $\mu = 31.94 \pm 0.15$~mag, respectively \citep{mould:2000}. 
 We note that the correction for Virgo infall \cite[see Appendix~A in][]{mould:2000} includes two components: the correction for the infall to Virgo plus the vector contribution due to the Local Group's peculiar velocity with respect to Virgo. 
We also note that in the Local Group the radial peculiar velocity dispersion is estimated to be $\sim 60$~km~s$^{-1}$ \citep[see for example][]{feldman:2003}, which accounts for about 25\% of the total error budget on $\mu$.


Soon after discovery, spectroscopic classifications of SN~2014ck were obtained independently at the Lick Observatory and under the Asiago Classification Program \citep{tomasella:2014}.
The earliest spectrum indicated it was a SN~Iax on the rise \citep{masi:2014}, resembling SN~2005hk \citep{phillips:2007}, SN~2008ha \citep{valenti:2009, foley:2009} and SN~2010ae \citep{stritz:2014}. 

Given the relatively small number of well-observed SNe~Iax in the literature and the early detection and classification, we initiated a follow-up campaign aimed to collect detailed optical and near-infrared (NIR) observations using several telescopes available to our collaboration. 

Basic information of SN~2014ck and its host galaxy are summarised in Table~\ref{info}. Following the discussion in Section~6, we adopt a $V$-band maximum estimate of ${\rm MJD} = 56845.6\pm0.1$, which we use as a reference epoch throughout this work.

\begin{table}\label{info}
\caption{Basic information on SN 2014ck and its host galaxy, UGC 12182. 
}
\begin{center}
\begin{tabular}{llll}
\hline 
Host galaxy  & UGC 12182 \\
Galaxy type     & Sbc \\
Heliocentric radial velocity   & $1490 \pm 19$ km s$^{-1}$\\
Distance modulus & $31.94 \pm 0.15$ mag\\
Galactic extinction A$_{V}$        & $1.26 \pm 0.15$ mag\\
Total extiction A$_{V}$               &  $\approx 1.5 \pm 0.3$ mag \\
                                   & \\
SN type & Iax\\
R.A.\ (J2000.0) & $22^{\rm h}45^{\rm m}38^{\rm s}\mathllap{.}88$\\                                   
Dec.\ (J2000.0) & $+73^\circ09'42\farcs7$ \\
Offset from nucleus & 4\farcs3 E, 0\farcs5 S\\
Estimated date of explosion (MJD) &  $56828.2^{+2.7}_{-4.5}$ \\
Date of first detection (MJD) & 56832.5 \\
Date of $V$-band maximum (MJD) &  $56845.6 \pm 0.1$ \\
$M_V$ at maximum & $-17.29 \pm 0.15$ mag\\ 
$M_B$ at maximum    & $-17.37 \pm 0.15$ mag\\
$L_{\rm bol}$ at maximum & $1.91 \times 10^{42}$ erg s$^{-1}$ \\                                   
\hline 
\end{tabular}
\end{center}
\label{info}
\end{table}%

\begin{figure}
\includegraphics[scale=.47,angle=0]{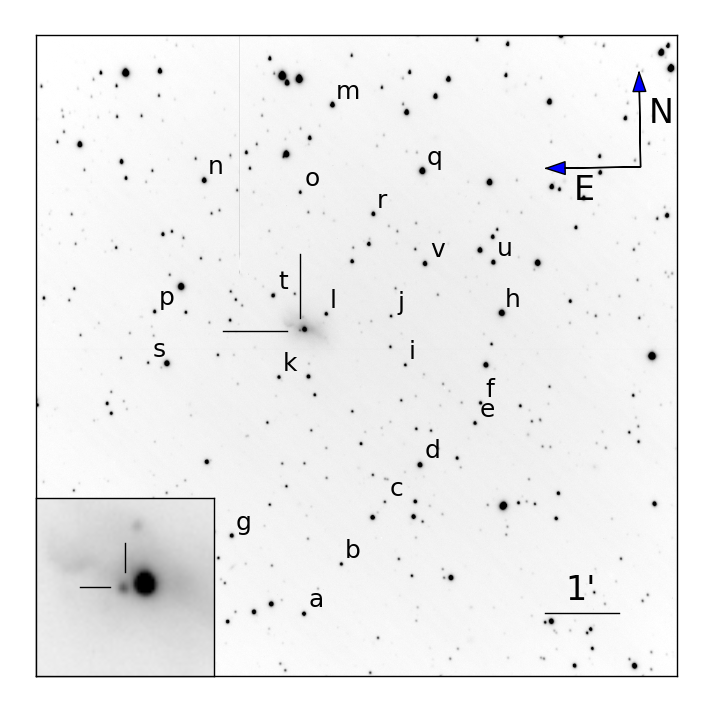}
\caption{UGC 12182 and SN~2014ck: $r$-band image taken on 2014 October 28.84 UT with the Copernico 1.82~m Telescope (Asiago), with an inset of the SN region on the bottom left. The  local sequence stars used for the calibration of non-photometric nights are indicated.}
\label{map1}
\end{figure}

\section{Hubble Space Telescope pre-explosion images}

\begin{figure}
\includegraphics[scale=.23,angle=0]{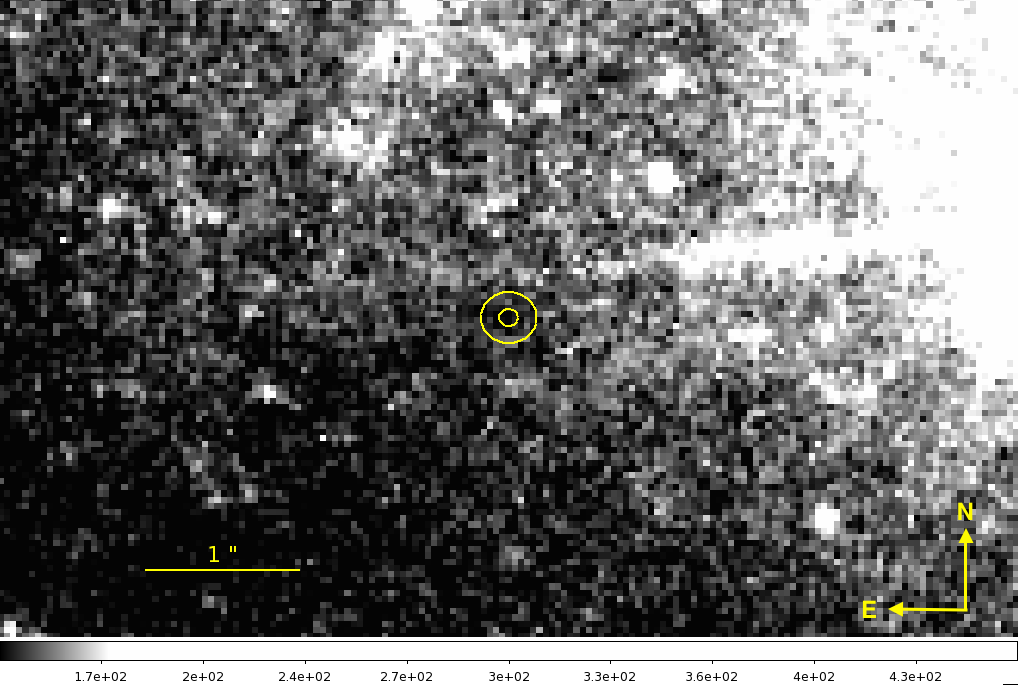}
\caption{{\it HST} pre-explosion image (F625W filter). The position of SN~2014ck is marked with ellipses. The outer ellipse corresponds to three times the uncertainty in the SN position.}
\label{map}
\end{figure}

Another transient, SN~2006fp \citep{puckett} was previously discovered in the host galaxy of SN~2014ck. 
The nature of this transient is unclear, but it was likely a SN~IIn or a SN impostor (i.e. the outburst of a luminous blue variable star), the latter being favoured by its spectral characteristics \citep{blondin}. 

{\sl HST} imaging was obtained of the host of SN~2006fp with the Ultraviolet-Visible (UVIS) Channel of the Wide Field Camera 3 (WFC3) (pixel scale $0\farcs04$ pix$^{-1}$). 
Images were taken on 2013 February 22 UT ({\sl HST} proposal ID 13029; PI: A.~Filippenko) with the F625W (roughly $r$) and F814W (roughly $I$) passbands.
The archival flat-fielded images ({\sc flt}) were retrieved  from the {\sl HST} MAST 
Archive\footnote{https://archive.stsci.edu/hst/} and re-reduced using the WFC3 UVIS CTE correction software\footnote{http://www.stsci.edu/hst/wfc3/}
and the {\sc AstroDrizzle} software from the {\sc DrizzlePac} package \citep{gonzaga:2012}.

Next the absolute astrometry was registered to match the ground-based $g$-band images which were obtained with the LCOGT 2.0~m Telescope (Haleakala, Hawaii, USA) on 2014 August 11.37 UT. Astrometric alignment was accomplished by fitting a second-order Legendre polynomial with the {\sc iraf}\footnote{{\sc iraf} is distributed by the National Optical Astronomy Observatory, which is operated by the Association of Universities for Research in Astronomy (AURA) under a cooperative agreement with the National Science Foundation.}
tasks {\sc geomap} and {\sc geoxytran}, measuring the position of 16 stars that were visible in both the LCOGT and HST frames. This yielded an astrometric precision of 0\farcs033 and 0\farcs022 in the east-west and north-south directions, respectively. 

The position of SN~2014ck in the LCOGT image was determined by fitting a Gaussian to the SN. We estimated the uncertainties in the position of the SN by running Markov chain Monte Carlo (MCMC) analysis using the {\sc emcee} Python package \citep{emcee}. We found that the uncertainty on the SN position was 50~milliarcseconds. 

Adding the astrometric solution and the positional uncertainties of the SN position in quadrature, we adopt a total uncertainty on the position in the F625W pre-explosion images to be 0\farcs06 and 0\farcs055 in the east-west and north-south directions, respectively (see Figure~\ref{map}).

We next used {\sc dolphot}\footnote{{\sc dolphot} is a stellar photometry package that was adapted from HSTphot \citep{dolphin00}.} to measure the photometry of all the stars in the pre-explosion images. The sky subtraction (in this case, the sky subtraction includes the diffuse contribution from the host galaxy) and PSF fits were done using the recommended parameters from the {\sc dolphot} manual. 

No source was detected within $3\sigma$ of the position of SN~2014ck. Using the detected sources from a $200\times200$ box centered around the SN position, we found $3\sigma$ limiting magnitudes of $m_{\rm F625W} > 26.95$ and $m_{\rm F814W} > 26.35$~ mag in the Vega system. Adopting a distance modulus $\mu = 31.94$~mag and reddening estimate $E(B-V)_{\rm tot} \approx 0.5$~mag (see Section~5), we obtain absolute luminosity limits $M_{\rm F625W} > -6.5$~mag. In passing, we note that there is no evidence of a stellar source at the position of SN~2006fp.  

The search for progenitor candidates in pre-explosion {\it HST} images has previously been performed for the Type~Iax SNe~2008ge \citep{foley:2010a}, 2012Z \citep{mccully:2014}, and 2014dt \citep{foley:2015}. 
At position of SN~2012Z, \cite{mccully:2014} detected a bright ($M_{\rm F435W}=-5.43 \pm 0.15$~mag, $M_{\rm F814W}=-5.24 \pm 0.16$~mag, i.e. $M_{V} \sim -5.3$~mag) blue source, which they interpret as a non-degenerate He-star companion to a C/O WD. The source associated with SN~2012Z is the only probable progenitor system detected in pre-explosion images of SNe~Iax and of any SN~Ia \citep[][and references therein]{li:2011}.
\cite{wang:2014} and \cite{liua:2015,liub:2015} performed binary evolution simulations indicating this could indeed explain the observed photometry. 
Yet the possibility that the source associated with SN~2012Z is a massive star cannot be entirely ruled out. 
Planned observations after the fading of the SN may help to finally distinguish between these progenitor models \citep{mccully:2014}.  
 
In the cases of SNe~2008ge and 2014dt, non-detections were reported and {\it HST} images 
were used by \cite{foley:2010a} and \cite{foley:2015} to place $3\sigma$ limits on the absolute magnitudes of the progenitors, corresponding to $M_{V} > -6.7$~mag  
and a relatively deep $M_{F450W} > -5.0$~mag, respectively. 
For SN~2014dt, \cite{foley:2015} excluded a massive star as the progenitor and suggested a C/O WD primary plus a He-star companion progenitor system, similar to SN~2012Z.  For both SN~2008ge \citep{foley:2010a} and SN~2014ck, the constraints on the luminosity of the undetected progenitor, a magnitude brighter than the SN~2012Z detection, rules out only the most-luminous Wolf-Rayet stars \citep[roughly corresponding to stars with initial masses of $60-65 M_\odot$,][]{crowther:2007}.

\section{Observation and data reduction}

\subsection{Photometry}

\begin{table*}
\caption{List of observing facilities employed for optical and NIR photometry.} \label{telescopes_phot}
\begin{tabular}{lllcc}
\hline 
Telescope$^1$                       & Instrument & Site & FoV  & Scale \\
                                         &                     &        & [arcmin$^{2}$] & [arcsec pix$^{-1}$]\\
\hline
\multicolumn{5}{c}{ \bf Optical facilities }\\
LCOGT                             & Spectral$^2$  & Haleakala, Hawaii (USA)                                            &    $10 \times 10$  & 0.30    \\
LCOGT                              & SBIG   & McDonald Observatory,  Texas (USA)                                            &    $16 \times 16$ & 0.47    \\
Copernico         & AFOSC      & Asiago, Mount Ekar (Italy) &   $8.8 \times 8.8$   &  0.48 \\
NOT                       & ALFOSC     & Roque de los Muchachos, La Palma, Canary Islands (Spain) & $ 6.4 \times 6.4 $ & 0.19 \\
TNG                      &  LRS &  Roque de los Muchachos, La Palma, Canary Islands (Spain) &$8.6 \times 8.6 $ & 0.25 \\
\multicolumn{5}{c}{ \bf NIR facilities }\\
NOT                  & NOTCam     & Roque de los Muchachos, La Palma, Canary Islands (Spain) & $4 \times 4 $ & 0.23\\ 
TNG                     & NICS    & Roque de los Muchachos, La Palma, Canary Islands (Spain) &$4.2 \times 4.2 $ & 0.25 \\ 
\hline 
\end{tabular}

$^1$ LCOGT = Las Cumbres Observatory Global Telescope Network\citep{brown:2013}; Copernico = INAF Osservatorio Astronomico di Padova 1.82~m Telescope (Mt.~Ekar, Asiago, Italy); NOT = 2.56~m Nordic Optical Telescope (La Palma, Spain); TNG = 3.58m Telescopio Nazionale Galileo (La Palma, Spain). 

$^2$ Spectral is a photometric camera mounted on the Faulkes Telelescopes of the LCOGT network. 
\end{table*}

\begin{figure}
\includegraphics[scale=.52, angle=0]{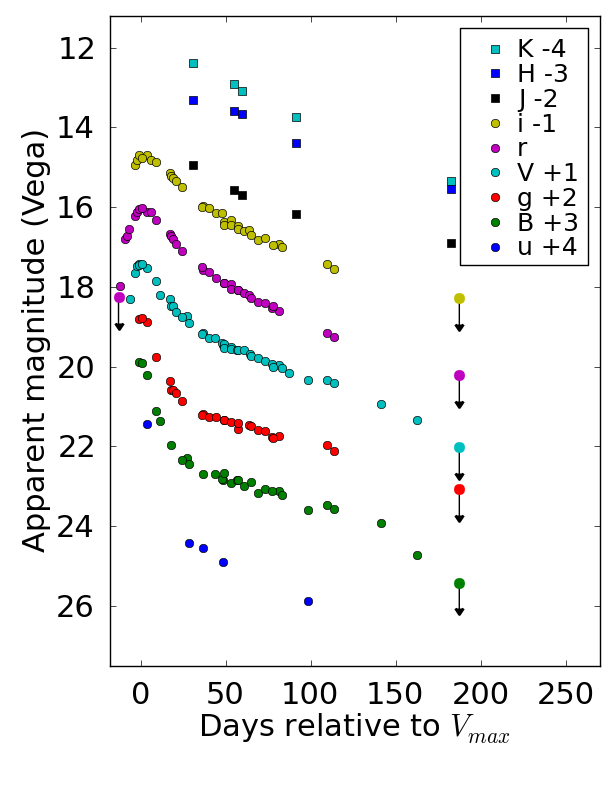}
\caption{Light curves of SN~2014ck in the $uBVgriJHK$ bands. Sloan $ugri$ AB magnitudes have been here plotted as Vega magnitudes for uniformity with $BVJHK$ bands, following Blanton \& Roweis 2007. For clarity, the light curves have been shifted vertically as indicated in the legend. The uncertainties for most data points are smaller than the plotted symbols. The last $BVgri$ photometric epoch is an upper limit. 
(A colour version of this figure is available in the online journal).} \label{lc}
\end{figure}

Optical ($uBVgri$) and NIR ($JHK$) imaging of SN~2014ck started a few days after discovery and continued over the course of about six months. The telescopes and their associated instruments used for the photometric campaign are listed in Table~\ref{telescopes_phot}.  

All frames were pre-processed using standard procedures in {\sc iraf} for bias subtraction and flat fielding. For the NIR exposures, sky background subtraction was also performed. Multiple exposures obtained in the same night were aligned and combined to increase the signal-to-noise ratio. 

Over the course of multiple photometric nights, \cite{landolt:1992} and Sloan Digital Sky Survey (SDSS)\footnote{http://www.sdss.org} standard fields were observed in order to calibrate a local sequence of stars in the field of UGC~12182 (see Table~\ref{sequence} and Figure~1). The local sequence was used to compute zero-points for non-photometric nights. In the NIR, stars from the 2MASS catalog were used for the calibration. 
We verified that photometry taken at similar phases but with different instrumentation were in excellent
agreement with each other, checking, for all bands, the RMS dispersion of the whole data set with respect to the dispersion of the sub-sets coming from each instrument. Thus no additional $S$-correction \citep{stritz:2002} was applied.

\begin{table*}
\caption{Magnitudes for the local sequence stars, as indicated in Figure~1, with associated errors in parentheses (Vega mag).}\label{sequence}
\begin{tabular}{cccccccc}
\hline 
ID &          R.A.\ (J2000.0) &          Dec.\ (J2000.0) &          $U$ &           $B$ &           $V$ &     $R$ &         $I$ \\
         &         & & [mag]&[mag]&[mag]&[mag]&[mag]\\
\hline 
a&22:45:38.708 & 73:05:53.33 &19.241 (0.024)& 18.137 (0.016)& 16.848 (0.023)& 16.047 (0.019)& 15.443 (0.013)\\
b&22:45:31.661 &73:06:33.20 &18.992 (0.026)& 18.566 (0.015)& 17.566 (0.019)& 16.918 (0.006)& 16.393 (0.013)\\
c&22:45:23.408 & 73:07:22.70 &19.537 (0.019)& 19.373 (0.008)& 18.496 (0.016)&         $-$             &$-$\\
d&22:45:16.752  &73:07:52.77 &18.649 (0.002)& 17.425 (0.008)& 16.042 (0.015)&          $-$            &14.471 (0.006)\\
e&22:45:06.405  &73:08:25.75 &18.402 (0.006)& 18.189 (0.002)& 17.295 (0.008)& 16.646 (0.002)& 16.119 (0.014)\\
f&22:45:05.321  &73:08:42.04 &18.833 (0.028)& 18.526 (0.003)& 17.568 (0.012)& 16.867 (0.004)& 16.325 (0.022)\\
g&22:45:51.990  &73:06:56.89 &17.739 (0.024)& 17.501 (0.012)& 16.631 (0.013)& 16.078 (0.004)& 15.570 (0.010)\\
h&22:45:01.180  &73:09:54.94 &18.027 (0.010)& 16.653 (0.004)& 15.310 (0.016)& 14.397 (0.021)& 13.565 (0.027)\\
i&22:45:19.245  &73:09:13.70 &19.239 (0.008)& 18.956 (0.010)& 17.986 (0.011)& 17.408 (0.028)& 16.747 (0.008)\\
j&22:45:21.834  &73:09:53.57 &19.537 (0.008)& 19.171 (0.012)& 18.166 (0.014)& 17.636 (0.018)& 16.975 (0.008)\\
k&22:45:42.885  &73:09:05.03 &18.547 (0.019)& 18.216 (0.008)& 17.381 (0.010)& 16.928 (0.014)& 16.365 (0.012)\\
l&22:45:33.912  &73:09:55.99 &19.115 (0.005)& 18.457 (0.010)& 17.398 (0.006)& 16.804 (0.011)& 16.144 (0.016)\\
m&22:45:32.298  &73:12:45.27& 18.003 (0.002)& 17.157 (0.005)& 15.966 (0.017)& 15.329 (0.012& 14.566 (0.006)\\
n&22:45:56.437  &73:11:45.34 &17.787 (0.017)& 17.024 (0.009)& 15.908 (0.010)& 15.322 (0.014)& 14.613 (0.013)\\
o&22:45:38.507  &73:11:34.89 &18.715 (0.008)& 18.475 (0.002)& 17.628 (0.026)& 17.171 (0.006)& 16.534 (0.002)\\
p&22:46:05.921  &73:09:59.35 &18.456 (0.005)& 18.117 (0.005)& 17.233 (0.019)& 16.754 (0.012)& 16.144 (0.008)\\
q&22:45:15.643  &73:11:50.85 &15.884 (0.008)& 15.661 (0.011)& 14.865 (0.012)& 14.406 (0.014)& 13.813 (0.016)\\
r&22:45:24.907  &73:11:16.49 &17.944 (0.017)& 17.647 (0.011)& 16.715 (0.012)& 16.195 (0.005)& 15.529 (0.016)\\
s&22:46:07.248  &73:09:17.79 &19.529 (0.001)& 19.220 (0.009)& 18.343 (0.011)& 17.887 (0.017)& 17.313 (0.009)\\
t&22:45:43.815  &73:10:11.37 &18.262 (0.002)& 17.925 (0.013)& 17.024 (0.018)&           $ -$              &$-$\\
u&22:45:02.577  &73:10:36.09 &17.689 (0.003)& 17.329 (0.007)& 16.410 (0.005)& 15.770 (0.005)& 15.171 (0.027)\\
v&22:45:15.320  &73:10:35.72 &17.378 (0.000)& 17.130 (0.005)& 16.199 (0.002)& 15.660 (0.004)& 15.066 (0.013)\\
\hline 
\end{tabular}
\end{table*}

All  photometry was performed via point spread function (PSF) fitting using the {\sc SNOoPY} package \citep{Cappellaro:2014}. SNOoPY is a collection of {\sc python} scripts calling standard {\sc iraf} tasks (through {\sc pyraf}) and specific data analysis tools such as {\sc sextractor} for source extraction and {\sc daophot} for PSF fitting. 
The sky background at the SN location is first estimated with a low-order
polynomial fit to data in the surrounding area. Then, the PSF model derived from isolated field stars is simultaneously fitted to the SN and any point source projected nearby (i.e. any star-like source within a radius of $\sim 5\times {\rm FWHM}$ from the SN).  The fitted sources are removed from the original images, an improved estimate of the local background derived and the PSF fitting procedure iterated. The residuals are visually inspected to validate the fit. 

An alternative approach for the measurement of transient magnitudes is template subtraction. The application of this technique requires the use of exposures of the field obtained before the SN explosion or after the SN has faded. The template images need to be in the same filter and have good signal-to-noise and seeing. Unfortunately, we could not find archival images suitable for use as templates, so only the PSF-fitting procedure was performed. On the contrary, for earlier epochs of LOSS/KAIT imaging, the pre-explosion image obtained on 2014 June 13 was used as a subtraction template (see Section~\ref{discovery}). 


Error estimates for the SN magnitudes are obtained through artificial star experiments in which a fake star with a similar magnitude to the SN is placed in the fit residual image at a position close to, but not coincident with, the SN location. 
The simulated image is processed through the same PSF fitting procedure and
the standard deviation of magnitudes of the fake stars is taken as an
estimate of the instrumental magnitude error, which is mainly due to the
uncertainty in the background fitting. This is combined (in quadrature) with the PSF fit error returned by {\sc daophot} and the propagated errors from the photometric calibration chain.

Johnson/Bessel and Sloan optical magnitudes of the SN and associated errors are listed in Tables~\ref{joh} and \ref{sloan}, respectively, while the NIR photometry is given in Table~\ref{nir_data}.  
Magnitudes are in the Vega system for the Johnson/Bessel filters and are close to the AB system (${\rm SDSS} = {\rm AB} - 0.02$~mag) for the Sloan filters.

The $uBVgriJHK$ light curves of SN~2014ck are plotted in Figure~\ref{lc}. Note
that since only a handful of $RIz$ epochs are available, we list their values in Tables~4 and 5 but do not plot them.

\subsection{Spectroscopy}

\begin{table*}
\caption{Journal of spectroscopic observations.} \label{telescope_spec}
\begin{tabular}{c c c c c c}
\hline
Date	  &MJD      &Phase$^1$	& Instrumental & Range & Resolution$^3$        \\
&     &[d]         & configuration$^2$& [\AA]  & [\AA]		 \\ 
\hline 
20140701 & 56839.58   & $-$6.0  &LCOGT+FLOYDS  &3200-10000&13\\       
20140702 & 56840.52   & $-$5.0  &LCOGT+FLOYDS  &3200-10000&13\\
20140703 & 56841.52   & $-$4.0  &LCOGT+FLOYDS  &3200-10000&13\\
20140703 & 56841.98   & $-$3.6  &Ekar+AFOSC+gm4 &3500-8200&24   \\
20140704 & 56842.57   & $-$3.0  &LCOGT+FLOYDS  &3200-10000&13\\
20140706 & 56844.52   & $-$1.0  &LCOGT+FLOYDS  &3200-10000&13   \\
20140706 & 56844.55   & $-$1.0  & Gemini-N+GNIRS & 9800-25000 & 4 \\
20140707 & 56845.50   & $-$0.1    & Gemini-N+GNIRS& 9800-25000 & 4 \\
20140709 & 56847.21   &  1.7       & NOT+ALFOSC+gm4   &3400-9000 &14  \\
20140710 & 56848.49   &  2.9       & LCOGT+FLOYDS  &3200-10000 &13   \\
20140711 & 56849.19   &  3.6       & TNG+LRS+LR-B        &3200-8000&10   \\
20140711 & 56849.45   &  3.9      & Gemini-N+GNIRS & 9800-25000 & 4 \\
20140712 & 56850.58   &  5.0       & LCOGT+FLOYDS  &3200-10000 &13  \\
20140718 & 56856.99   & 11.4     &  Ekar+AFOSC+gm4+VPH6 &3500-9300 &24  \\
20140724 & 56862.20   & 16.6     &   NOT+ALFOSC+gm4   &3400-9000 &14  \\
20140725 & 56863.58   & 18.0     &  LCOGT+FLOYDS  &3200-10000 &13  \\
20140726 & 56864.51   & 19.0      & Gemini-N+GNIRS & 9800-25000 & 4 \\
20140727 & 56865.56   & 20.0     &   LCOGT+FLOYDS  &3200-10000 &13   \\
20140728 & 56866.54   & 21.0    &    LCOGT+FLOYDS  &3200-10000 &13 \\
20140731 & 56869.28   & 23.7      & Gemini-N+GNIRS & 9800-25000 & 4 \\
20140801 & 56870.18   & 24.6    &    NOT+ALFOSC+gm4   &3400-9000 &14  \\
20140805 & 56874.50   & 28.9    &   LCOGT+FLOYDS  &3200-10000 &13 \\
20140807 & 56876.39   & 30.8      & Gemini-N+GNIRS & 9800-25000 & 4 \\
20140812 & 56881.98   & 36.4            & TNG+LRS+LR-B+LR-R & 3500-10000& 10\\
20140815 & 56884.54   & 39.0    &    LCOGT+FLOYDS  &3200-10000 &13\\
20140823 & 56892.36   & 46.8    &  LCOGT+FLOYDS  &3200-10000 &13 \\
20140825 & 56894.00  &  48.4   &  NOT+ALFOSC+gm4 &3400-9000 &14 \\
20140831 & 56900.94  &  55.4    & TNG+NICS+IJ+HK   &  9000-17500& 6\\
20140926 & 56926.94  &  81.4   & Ekar+AFOSC+gm4+VPH6 &3500-9300 &24  \\
20141025 & 56955.94  &  110.4 & TNG+LRS+LR-B+LR-R & 3500-10000& 10\\
20141220 & 57011.89   & 166.3   & GTC+OSIRIS+R300B   & 3500-9000 & 16 \\
\hline 
\end{tabular}

$^1$ The phase is relative to the adopted epoch of the $V$-band maximum, ${\rm MJD} = 56845.6 \pm 0.1$.

$^2$ NOT = 2.56~m Nordic Optical Telescope (La Palma, Spain); Ekar = Copernico 1.82~m Telescope (Mt.~Ekar, Asiago, Italy); TNG = 3.58~m Telescopio Nazionale Galileo (La Palma, Spain); LCOGT = LCOGT 2.0~m Telescope (Haleakala, Hawaii, USA); Gemini-N = 8.1~m Telescope (Hilo, Hawaii, USA); GTC = 10.4~m Gran Telescopio Canarias (La Palma, Spain). 

$^3$ The resolution is estimated from the FWHM of the night sky lines. 
\end{table*}

A sequence of 24 low-resolution visual-wavelength spectra for SN~2014ck were obtained extending from  $-6.0$~d to $+166.3$~d relative to the epoch of $V$-band maximum. Seven epochs of NIR spectra were also taken extending from $-1.0$~d to $+55$~d.
A summary of all spectroscopic observations is provided in Table~\ref{telescope_spec}. 

Optical spectra were reduced using standard {\sc iraf} tasks. After bias and flat-field correction, the SN spectrum was extracted and calibrated in wavelength through a comparison to arc lamp spectra. The flux calibration was derived from observations of spectrophotometric standard stars obtained, when possible, on the same night as the SN. All the flux-calibrated spectra were verified against photometry and, when necessary, a correction factor was applied. Corrections for the telluric absorption bands were derived using the spectrophotometric standard star spectra. 
In some cases a non-perfect removal can affect the SN features that overlap with the strongest atmospheric absorptions, 
in particular with the telluric O$_{2}$ $A$-band at $7590-7650$~\AA\/ and the H$_{2}$O, CO$_{2}$, CH$_{4}$ bands in NIR spectra (their positions are marked in Figures~8, 9, 10, 11, 13, 14 and 15 with the $\oplus$ symbol and, for the strongest ones, with vertical grey bands).


The NIR spectra obtained with GNIRS attached to the Gemini North telescope were reduced using the {\sc gnirs} Gemini {\sc iraf} package \citep[see][for details]{hsiao:2013}. 
The TNG spectrum obtained with the Near Infrared Camera Spectrograph (NICS) was reduced using standard {\sc iraf} packages. 
In brief, following the standard infrared technique, each night several pairs of spectra were taken at different positions along the slit, and consecutive pairs were subtracted from each other in order to remove the sky background.
The subtracted images were aligned to match the stellar profile and added together. Finally, the source spectrum was extracted from the combined images. Wavelength calibration, telluric correction and flux calibration were done in the standard manner. Lastly, spectra were corrected to match the broadband photometry. 


\section{Galactic and host reddening}\label{extinction}

The Galactic extinction in the direction of UGC~12182, as derived 
from the \cite{schlafly:2011} recalibration of the \cite{schlegel:1998} infrared-based dust map, is $E(B-V)_{\rm G} = 0.40 \pm 0.05$~mag (via NED), which corresponds to a Galactic extinction $A_V = 1.26 \pm 0.15$~mag when adopting a standard 
$R_V = 3.1$ reddening law \citep{cardelli:1989}. 

The extinction within the host galaxy is more uncertain. A standard approach for SNe Ia is to measure the colour excess by comparing the SN colour with that of an unreddened SN template. However, the comparative study of the $B-V$, $V-R$ and $V-I$ colour curves for a sample of 
SNe~Iax presented by \cite{foley:2013} shows significant scatter that does not improve after reddening corrections. So far, it is unclear if these objects have similar intrinsic colours or not. 
High-dispersion observations of Na~I~D $\lambda\lambda$5890, 5896 are used as an independent means of probing dust extinction to extragalactic sources \citep{poznanski:2012}. However, for medium- to low-resolution spectra, when the doublet is blended, there is a large scatter in the data \citep{turatto:2003,poznanski:2011} and the correlation has less predictive power. Moreover, \cite{phillips:2013} showed that the column density and/or equivalent width (EW) of the Na~I~D lines are, in general, unreliable indicators of the extragalactic dust extinction suffered by SNe~Ia. 
The exception to this statement is that weak or undetectable Na~I absorption appears to be consistent with little or no extinction. 
With this caveat in mind, the earlier spectra with the highest signal-to-noise ratio were selected from our spectral sequence and used to measure an average EW for the Galactic Na~I~D of $2.8 \pm 0.3$~\AA. Following \cite{turatto:2003}, a lower limit for the colour excess within the Milky Way of $E(B-V)_{\rm G} = 0.44 \pm 0.05$~mag was found, in fair agreement with $E(B-V)_{\rm G} = 0.40 \pm 0.05$~mag obtained from the infrared maps of the galactic dust distribution. 
Only a weak absorption line, ${\rm EW} \lesssim 0.3$~\AA, may be attributed to the host Na~I~D. This is consistent with little extinction in the host galaxy   
($E(B-V)_{\rm host} \lesssim 0.05$~mag). Therefore, the total colour excess of the SN is estimated to be 
$E(B-V)_{\rm tot} \approx 0.5 \pm 0.1$~mag (i.e. $A_V \approx 1.5 \pm 0.3$~mag).

\section{Photometric evolution}\label{photo}

\begin{table*}
\caption{Optical photometry of SN~2014ck in the Johnson/Cousins $UBVRI$ filters (Vega mag), with associated errors in parentheses.} \label{joh}
\begin{tabular}{cccccccc}
\hline 
Date & MJD & $U$ & $B$ & $V$ & $R$  & $I$ & Instrument\\ 
         &         & [mag]&[mag]&[mag]&[mag]&[mag]&\\
\hline 
20140630 & 56838.83 & $-$ & $-$  & $-$  &  16.1 (0.5) & $-$  & Masi$^{1}$  \\ 
20140701 & 56839.38 & $-$ & $-$  &  17.34 (0.23) & $-$  & $-$  & Brimacombe$^{1,2}$  \\ 
20140701 & 56839.91 & $-$ & $-$  &$-$   &  16.1 (0.5) &$-$   & James$^{1}$  \\ 
20140704 & 56842.01 & $-$ &  $-$ &  16.64 (0.22) &$-$   &$-$& AFOSC  \\ 
20140705 & 56843.57 & $-$ & $-$  &  16.49 (0.20) & $-$  & $-$  & Spectral$^3$  \\ 
20140706 & 56844.42 & $-$ &  16.89 (0.06) &  16.44 (0.04) & $-$  & $-$  & SBIG  \\ 
20140706 & 56844.60 & $-$ & $-$  &  16.44 (0.22) & $-$  & $-$  & Spectral$^3$  \\ 
20140708 & 56846.30 & $-$ &  16.91 (0.03) &  16.43 (0.03) & $-$  & $-$  & SBIG  \\ 
20140711 & 56849.10 & $-$ &  17.22 (0.02) &  16.53 (0.01) & $-$  &  $-$ & LRS  \\ 
20140716 & 56854.44 & $-$ &  18.11 (0.04) &  16.85 (0.03) & $-$  & $-$  & Spectral$^3$  \\ 
20140719 & 56857.07 & $-$ &  18.35 (0.05) &  17.20 (0.07) &$-$&$-$   & AFOSC  \\ 
20140724 & 56862.58 & $-$ & $-$  &  17.31 (0.15) &$-$&$-$   & Spectral$^3$  \\ 
20140725 & 56863.45 & $-$ &  18.97 (0.04) &  17.47 (0.04) &  $-$ & $-$  & Spectral$^3$  \\ 
20140726 & 56864.31 & $-$ & $-$  &  17.49 (0.39) & $-$  & $-$  & SBIG  \\ 
20140728 & 56866.49 & $-$ & $-$  &  17.62 (0.17) &  $-$ & $-$  & Spectral$^3$  \\ 
20140731 & 56869.54 & $-$ &  19.35 (0.05) &  17.76 (0.04) &$-$&$-$   & Spectral$^3$  \\ 
20140803 & 56872.97 & $-$ &  19.29 (0.06) &  17.72 (0.03) & $-$  &  $-$ & AFOSC  \\ 
20140805 & 56874.13 & 20.18 (0.06) &  19.45 (0.02) &  17.91 (0.01) &  17.27 (0.08) &  16.64 (0.01) & ALFOSC \\ 
20140812 & 56881.36 & $-$ & $-$  &  18.19 (0.23) &$-$   &$-$& Spectral$^3$  \\ 
20140813 & 56882.01 & $-$ &  19.69 (0.03) &  18.16 (0.02) &$-$   &$-$   & LRS  \\ 
20140816 & 56885.36 & $-$ & $-$  &  18.29 (0.18) &$-$&$-$& Spectral$^3$  \\ 
20140820 & 56889.46 & $-$ &  19.69 (0.15) &  18.29 (0.03) & $-$  & $-$  & Spectral$^3$  \\ 
20140824 & 56893.42 & $-$ &  19.82 (0.14) &  18.40 (0.03) & $-$  &$-$   & Spectral$^3$  \\ 
20140824 & 56893.97 & 20.95 (0.07) &  19.84 (0.02) &  18.45 (0.03) &  17.80 (0.02) &  17.16 (0.02) & ALFOSC  \\ 
20140825 & 56894.36 & $-$ &  19.67 (0.06) &  18.43 (0.04) & $-$  & $-$  & Spectral$^3$  \\ 
20140825 & 56894.44 & $-$ & $-$  &  18.53 (0.21) & $-$  &$-$   & SBIG  \\ 
20140829 & 56898.36 & $-$ &  19.92 (0.15) &  18.52 (0.04) & $-$  &$-$   & Spectral$^3$  \\ 
20140829 & 56898.47 & $-$ & $-$  &  18.56 (0.20) & $-$  & $-$  & SBIG  \\ 
20140902 & 56902.26 & $-$ &  19.84 (0.13) &  18.58 (0.04) &$-$& $-$  & SBIG  \\ 
20140902 & 56902.53 & $-$ &  19.84 (0.05) &  18.58 (0.04) &$-$   &$-$& Spectral$^3$  \\ 
20140906 & 56906.32 & $-$ &  19.99 (0.07) &  18.59 (0.05) & $-$  & $-$  & Spectral$^3$  \\ 
20140909 & 56909.44 & $-$ & $-$  &  18.67 (0.09) & $-$  & $-$  & SBIG  \\ 
20140910 & 56910.30 & $-$ &  19.89 (0.10) &  18.74 (0.05) & $-$  &$-$   & Spectral$^3$  \\ 
20140914 & 56914.44 & $-$ &  20.17 (0.07) &  18.79 (0.04) & $-$  & $-$  & Spectral$^3$  \\  
20140918 & 56918.50 & $-$ &  20.06 (0.05) &  18.87 (0.03) & $-$  &$-$& SBIG  \\ 
20140922 & 56922.39 & $-$ &  20.13 (0.04) &  18.93 (0.04) & $-$  &$-$   & Spectral$^3$  \\ 
20140923 & 56923.39 & $-$ & $-$  &  19.00 (0.04) & $-$  &$-$   & Spectral$^3$  \\ 
20140927 & 56927.03 & $-$ &  20.11 (0.18) &  18.95 (0.21) &$-$& $-$  & AFOSC  \\ 
20140928 & 56928.32 & $-$ &  20.21 (0.18) &  19.03 (0.11) & $-$  &$-$& Spectral$^3$  \\ 
20141013 & 56943.83 & 21.94 (0.18) &  20.58 (0.03) &  19.32 (0.02) &  18.86 (0.06) &  18.04 (0.02) & ALFOSC  \\ 
20141024 & 56954.83 & $-$ &  20.47 (0.19) &  19.34 (0.21) &$-$&$-$& AFOSC  \\ 
20141028 & 56958.82 & $-$ &  20.58 (0.21) &  19.41 (0.21) & $-$  & $-$  & AFOSC  \\ 
20141125 & 56986.82 & $-$ &  20.93 (0.07) &  19.93 (0.10) &  19.30 (0.10) &  18.68 (0.15) & ALFOSC  \\ 
20141216 & 57007.85 & $-$ &  21.71 (0.08) &  20.34 (0.05) &  19.89 (0.06) &  18.76 (0.05) & ALFOSC  \\ 
20150110 & 57032.75 & $-$ &  22.41 (0.52) &  21.52 (0.56) & $-$  & $-$  & AFOSC  \\ 
\hline 
\end{tabular}

Notes: $^{1}$ From IAU CBET 3949 (Masi et al. 2014); $^{2}$ $V$ for reference; $^3$ ``Spectral'' is a photometric camera mounted 
on the Faulkes Telescopes of the LCOGT network. 
\end{table*}

\begin{table*}
\caption{Optical photometry of SN~2014ck in the Sloan $ugriz$ filters (AB mag), with associated errors in parentheses.} \label{sloan}
\begin{tabular}{cccccccc}
\hline 
Date & MJD & $u$ & $g$ & $r$ & $i$  & $z$ & Instrument\\ 
         &         & [mag]&[mag]&[mag]&[mag]&[mag]&\\
 \hline 
20140624 & 56832.46 & $-$ &$-$   &  18.41 (2.00) & $-$           &$-$   & KAIT$^{1,2}$ \\
20140625 & 56833.49 & $-$ &$-$   &  18.15 (0.44) & $-$           &$-$   & KAIT$^{1,3}$ \\
20140628 & 56836.49 & $-$ &$-$   &  16.96 (0.36) & $-$           &$-$   & KAIT$^{1}$ \\
20140629 & 56837.47 & $-$ &$-$   &  16.89 (0.11) & $-$           &$-$   & KAIT$^{1}$ \\
20140630 & 56838.41 & $-$ &$-$   &  16.71 (0.27) & $-$           &$-$   & KAIT$^{1}$ \\
20140704 & 56842.01 & $-$ &$-$   &  16.37 (0.02) &  16.32 (0.02) &$-$   & AFOSC  \\ 
20140705 & 56843.57 &$-$& $-$  &  16.32 (0.05) &  16.19 (0.06) & $-$  & Spectral$^5$  \\ 
20140706 & 56844.61 & $-$ &  16.73 (0.10) &  16.20 (0.08) &  16.09 (0.11) &$-$   & Spectral$^5$  \\ 
20140708 & 56846.32 &$-$  &  16.70 (0.06) &  16.19 (0.04) &  16.14 (0.05) & $-$  & SBIG  \\ 
20140711 & 56849.18 & 18.35 (0.02) &  16.81 (0.09) &  16.28 (0.03) &  16.06 (0.01) & $-$  & LRS  \\ 
20140713 & 56851.53 &$-$  & $-$  &  16.27 (0.03) &  16.18 (0.04) & $-$  & Spectral$^5$  \\ 
20140716 & 56854.44 & $-$ &  17.67 (0.04) &  16.48 (0.06) &  16.24 (0.05) &$-$& Spectral$^5$  \\ 
20140719 & 56857.07 & 20.20 (0.14) &  17.85 (0.03) &  16.60 (0.04) &  16.10 (0.03) &  17.70 (0.06) & AFOSC  \\ 
20140724 & 56862.58 &  $-$&  18.27 (0.05) &  16.83 (0.09) &  16.51 (0.11) & $-$  & Spectral$^5$  \\ 
20140725 & 56863.46 & $-$ &  18.49 (0.06) &  16.89 (0.04) &  16.59 (0.08) &  $-$ & Spectral$^5$  \\ 
20140726 & 56864.32 & $-$ &  18.50 (0.07) &  16.95 (0.06) &  16.64 (0.07) &  $-$ & SBIG  \\ 
20140728 & 56866.50 & $-$ &  18.57 (0.06) &  17.08 (0.05) &  16.70 (0.08) & $-$  & Spectral$^5$  \\ 
20140731 & 56869.54 & $-$ &  18.78 (0.05) &  17.26 (0.06) &  16.86 (0.09) &  $-$ & Spectral$^5$  \\ 
20140803 & 56872.97 & 21.08 (0.19) &  18.72 (0.04) &  17.34 (0.02) &  17.01 (0.02) &$-$& AFOSC  \\
20140805 & 56874.13 & 21.03 (0.06)  &$-$  & $-$  &$-$   & $-$  & ALFOSC$^{4}$  \\ 
20140812 & 56881.38 &  $-$&  19.13 (0.08) &  17.67 (0.05) &  17.37 (0.07) & $-$  & Spectral$^5$  \\ 
20140813 & 56882.01 & 21.46 (0.21) &  19.10 (0.09) &  17.75 (0.11) &  17.34 (0.13) &$-$& LRS  \\ 
20140816 & 56885.36 & $-$ &  19.19 (0.04) &  17.79 (0.06) &  17.39 (0.07) & $-$  & Spectral$^5$  \\ 
20140820 & 56889.47 & $-$ &  19.18 (0.05) &  17.93 (0.05) &  17.53 (0.08) &$-$& Spectral$^5$  \\ 
20140824 & 56893.43 & $-$ & $-$  &$-$   &  17.52 (0.34) & $-$  & Spectral$^5$  \\ 
20140824 & 56893.97 & 21.81 (0.07) & $-$  &$-$& $-$  & $-$  & ALFOSC$^{4}$  \\ 
20140825 & 56894.36 & $-$ &  19.27 (0.06) &  18.07 (0.05) &  17.75 (0.08) & $-$  & Spectral$^5$  \\ 
20140825 & 56894.46 &  $-$&  19.26 (0.11) &  18.06 (0.07) &  17.82 (0.07) &$-$& SBIG  \\ 
20140829 & 56898.38 & $-$ &$-$   &  18.08 (0.17) &  17.69 (0.08) & $-$  & Spectral$^5$  \\ 
20140829 & 56898.45 &  $-$&  19.32 (0.07) &  18.21 (0.06) &  17.82 (0.09) & $-$  & SBIG  \\ 
20140902 & 56902.45 &  $-$&  19.49 (0.11) &  18.23 (0.06) &  17.85 (0.08) &  $-$ & SBIG  \\ 
20140902 & 56902.54 &$-$  &  19.34 (0.07) &  18.25 (0.07) &  17.91 (0.08) &  $-$ & Spectral$^5$  \\ 
20140906 & 56906.33 &  $-$& $-$  &  18.32 (0.06) &  17.96 (0.10) &$-$& Spectral$^5$  \\ 
20140909 & 56909.42 &  $-$&  19.37 (0.15) &  18.36 (0.08) &  17.95 (0.20) &  $-$ & SBIG  \\ 
20140910 & 56910.32 &  $-$&  19.42 (0.06) &  18.44 (0.07) &  18.06 (0.04) &  $-$ & Spectral$^5$  \\ 
20140914 & 56914.46 &  $-$&  19.52 (0.07) &  18.53 (0.05) &  18.18 (0.10) &  $-$ & Spectral$^5$  \\ 
20140918 & 56918.51 &  $-$&  19.54 (0.07) &  18.55 (0.06) &  18.14 (0.10) &   $-$& Spectral$^5$  \\ 
20140922 & 56922.40 &  $-$&  19.68 (0.07) &  18.70 (0.08) & $-$  &$-$& Spectral$^5$  \\ 
20140923 & 56923.40 &  $-$&  19.70 (0.07) &  18.64 (0.16) &  18.31 (0.17) & $-$  & Spectral$^5$  \\ 
20140927 & 56927.04 &  $-$&  19.67 (0.19) &  18.77 (0.17) &  18.30 (0.04) &  19.33 (0.13) & AFOSC  \\ 
20141013 & 56943.83 & 22.79 (0.18) &$-$   & $-$  & $-$  & $-$  & ALFOSC$^{4}$  \\ 
20141024 & 56954.83 &  $-$&  19.87 (0.17) &  19.32 (0.19) &  18.80 (0.18) & $-$  & AFOSC  \\ 
20141028 & 56958.83 &  $-$&  20.03 (0.10) &  19.42 (0.10) &  18.93 (0.08) &  $-$ & AFOSC  \\ 
20150110 & 57032.77 &  $-$&  22.80 (0.36) &  22.07 (0.45) &  21.64 (0.47) &  $-$ & AFOSC  \\ 
\hline 
\end{tabular}

Notes: $^{1}$ LOSS/KAIT unfiltered images calibrated to $r$-band; $^{2}$ upper limit; $^{3}$ marginal detection; $^{4}$ $U$-band magnitude converted into $u$-band following \cite{chonis:2008}; $^5$ ``Spectral'' is a photometric camera mounted 
on the Faulkes Telescopes of the LCOGT network.

\end{table*}

\begin{table*}
\caption{Near-infrared photometry of SN~2014ck, with associated errors in parentheses.} \label{nir_data}
\begin{tabular}{cccccc}
\hline 
Date & MJD & $J$ & $H$ & $K$  & Instrument\\
        &           &[mag]&[mag]&[mag]&\\ 
\hline 
20140807 & 56876.08 & 16.94 (0.03) &  16.30 (0.01) &  16.39 (0.03) & NOTCam  \\ 
20140831 & 56900.08 & 17.58 (0.29) &  16.59 (0.17) &  16.91 (0.26) & NICS  \\ 
20140905 & 56905.03 & 17.70 (0.03) &  16.67 (0.05) &  17.08 (0.05) & NOTCam  \\ 
20141007 & 56937.07 & 18.16 (0.04) &  17.39 (0.07) &  17.75 (0.05) & NOTCam  \\ 
20150105 & 57027.83 & 18.89 (0.04) &  18.55 (0.09) &  19.34 (0.26) & NOTCam  \\ 
\hline 
\end{tabular}
\end{table*}

\subsection{Broad band photometry}\label{photo_evol}

\begin{figure}
\includegraphics[scale=.38,angle=0]{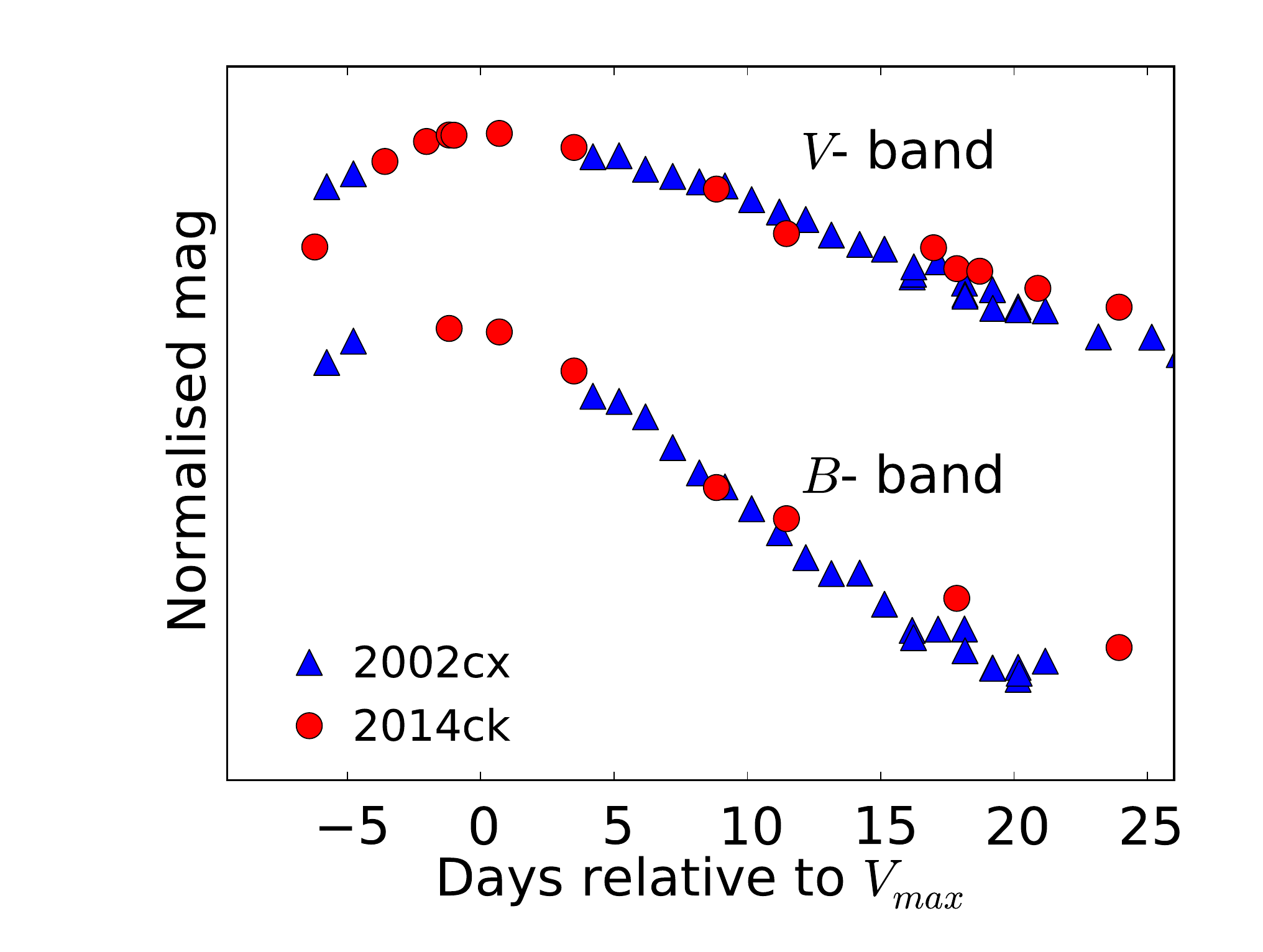}
\caption{Comparison of normalised (to maximum magnitude) $B$- and $V$-band light curves of SNe~2002cx and 2014ck. (A colour version of this figure is available in the online journal).} 
\label{lc2}
\end{figure}

\begin{figure}
\centering
\includegraphics[scale=.5]{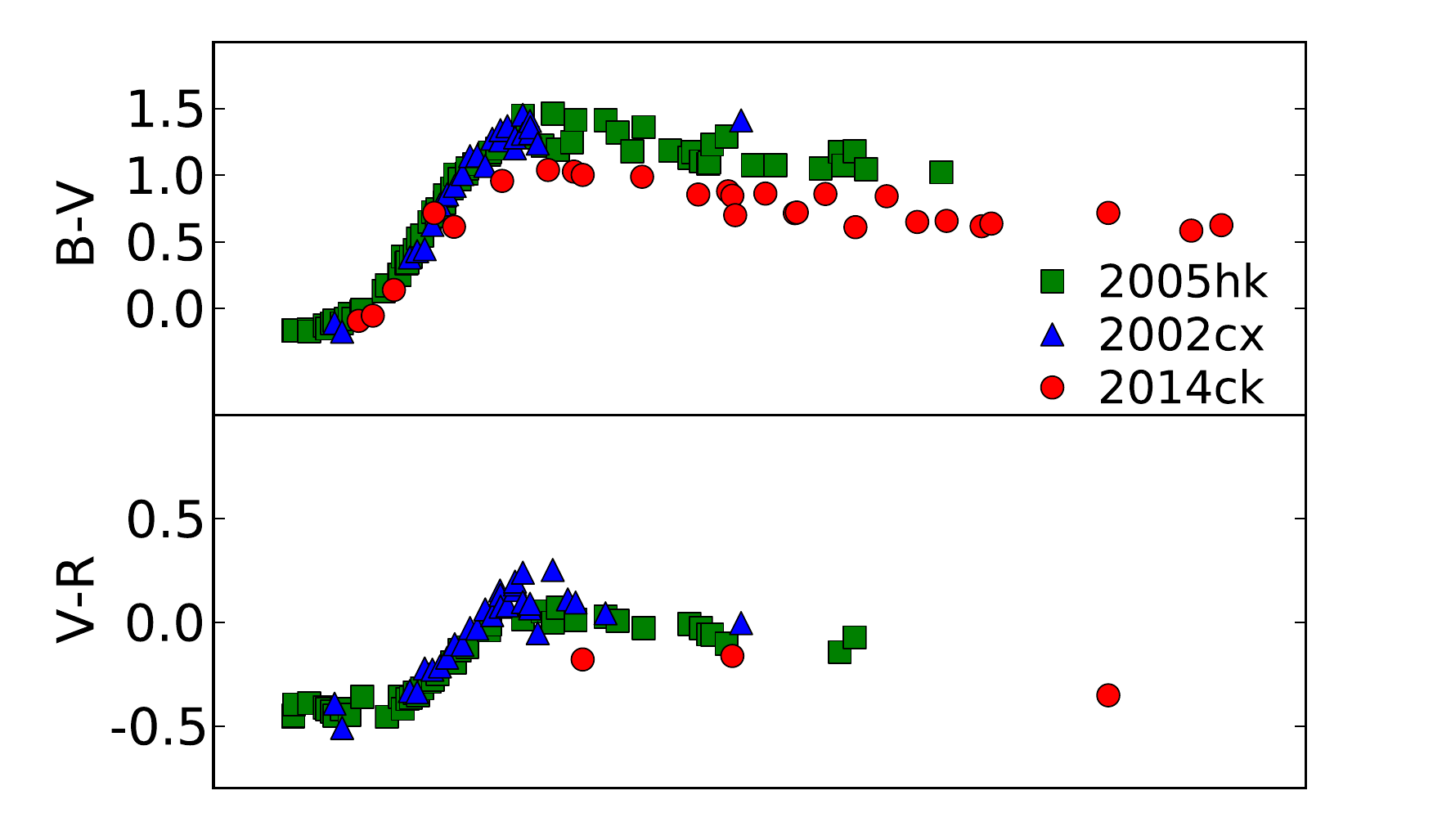}
\includegraphics[scale=.5]{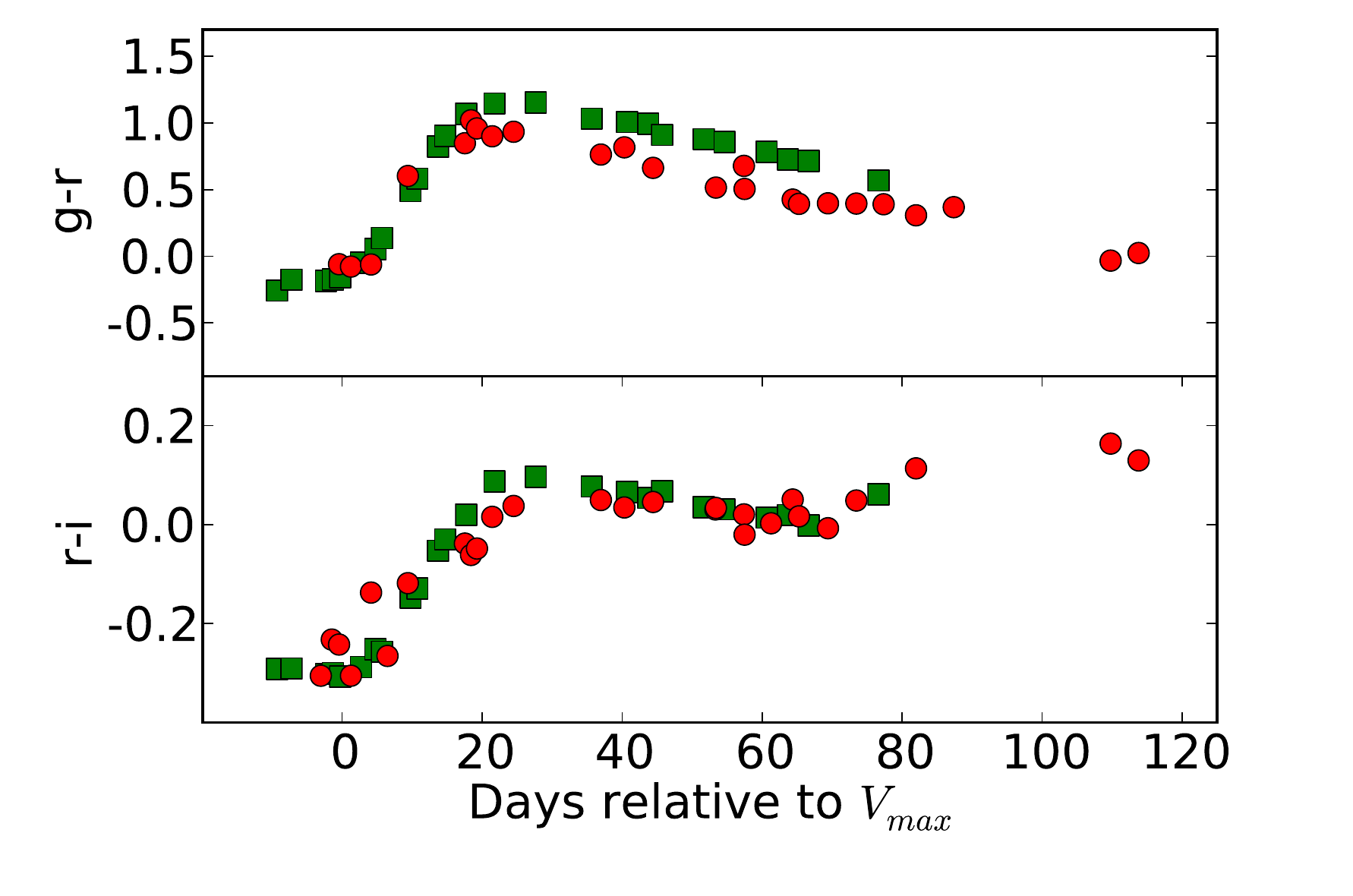}
\caption{From top to bottom: $B - V$ and $V - R$ (Vega mag), $g - r$ and $r - i$ (AB mag) extinction-corrected colours of SN~2014ck compared with those of SN~2002cx (blue triangles) and SN~2005hk (green squares). (A colour version of this figure is available in the online journal). }
\label{col}
\end{figure}

The photometric evolution of SN~2014ck shown in Figure~\ref{lc} is well sampled in the optical bands (except the pre-maximum evolution in $u$- and $B$-bands), while only a handful of NIR measurements were obtained. The light curves are characterised by a rise to maximum and a subsequent decline that is slower at longer wavelengths (e.g. in the $r$- and  $i$-band). Moreover, as already noted for other SNe~Iax \citep{li:2003,foley:2013,stritz:2015}, the NIR bands show no evidence of a secondary maximum, characteristic of normal SNe~Ia \citep{hamuya:1996,hamuyb:1996}. 
By using of a low-order polynomial fit to the optical light curves around maximum,
an estimate of the magnitude and epoch of maximum light for each band was obtained. SN~2014ck reached peak brightness $M_B=-17.37 \pm 0.15$~mag and $M_V=-17.29 \pm 0.15$~mag. All the absolute magnitudes are listed in Table~\ref{fit} along with their associated uncertainties estimated from the dispersion around the polynomial fit. Whereas the $gVri$ light curves are well sampled around maximum, the $B$-band light curve is already declining from the first $B$ point. Consequently, the time of $B$-band maximum might be ill constrained.

With best-fit peak apparent magnitudes in hand, absolute magnitudes were also computed, with associated uncertainties  obtained by adding in quadrature the errors of the fit to the peak apparent magnitudes and the errors in the adopted extinction and distance. 

Finally, our polynomial fits also provide a measure of the decline-rate parameter $\Delta m_{15}$, the magnitude drop from the epoch of maximum brightness to 15 days later. 
In the case of normal SNe~Ia, $\Delta m_{15}$ is known to correlate with luminosity \citep{phillips:1993}.
 

Examining the results of the polynomial fits, we find that maximum light is reached earlier in the blue bands compared to the red bands, with a delay of $\sim 4$ days between the epochs of $B$- and $i$-band maximum.
Furthermore, the blue bands have faster decline rates, with $\Delta m_{15} (B) = 1.76 \pm 0.15$~mag and $\Delta m_{15} (i) = 0.39 \pm 0.15$~mag. Both of these characteristics are common to all SNe~Iax \citep{foley:2013,stritz:2014,stritz:2015}.

As revealed from the comparison in Figure~\ref{lc2} of the $B$- and $V$-band light curves of SN~2002cx and SN~2014ck, the two objects show nearly identical evolution.
Moreover, as shown in Figure~\ref{bol}, the two objects reached the same peak bolometric luminosity. 
The decline rates of the two objects are also nearly identical: $\Delta m_{15}(B) = 1.76 \pm 0.15$~mag for SN~2014ck vs.\ $\Delta m_{15}(B) = 1.7 \pm 0.1$~mag for SN~2002cx. These values are significantly slower than the decline rate of the faint and fast SN~2008ha \citep[$\Delta m_{15}(B) = 2.03 \pm 0.20$~mag]{valenti:2009} and SN~2010ae \citep[$\Delta m_{15}(B) = 2.43 \pm 0.11$~mag]{stritz:2014}. In conclusion, SN~2014ck follows the general trend for SNe~Iax (and SNe~Ia, in general): more luminous objects tend to have broader light curves \citep{foley:2013}. 

Various optical-band colour curves of SN~2014ck, corrected for reddening, are plotted in Figure~\ref{col}. At early phases the colours are blue. As the SN evolves, the colours change towards the red, reaching a maximum value around three weeks past maximum. 
Subsequently, the colours slowly evolve back towards the blue.
Inspecting the colour curves of SN~2014ck compared to those of the Type~Iax SN~2002cx \citep{li:2003} and SN~2005hk \citep{phillips:2007,stritz:2015}, we note similar evolution, with SN~2014ck appearing marginally bluer over all epochs.

\subsection{Explosion date and rise time estimates}\label{rise}

\begin{figure}
\includegraphics[scale=.4, angle=0]{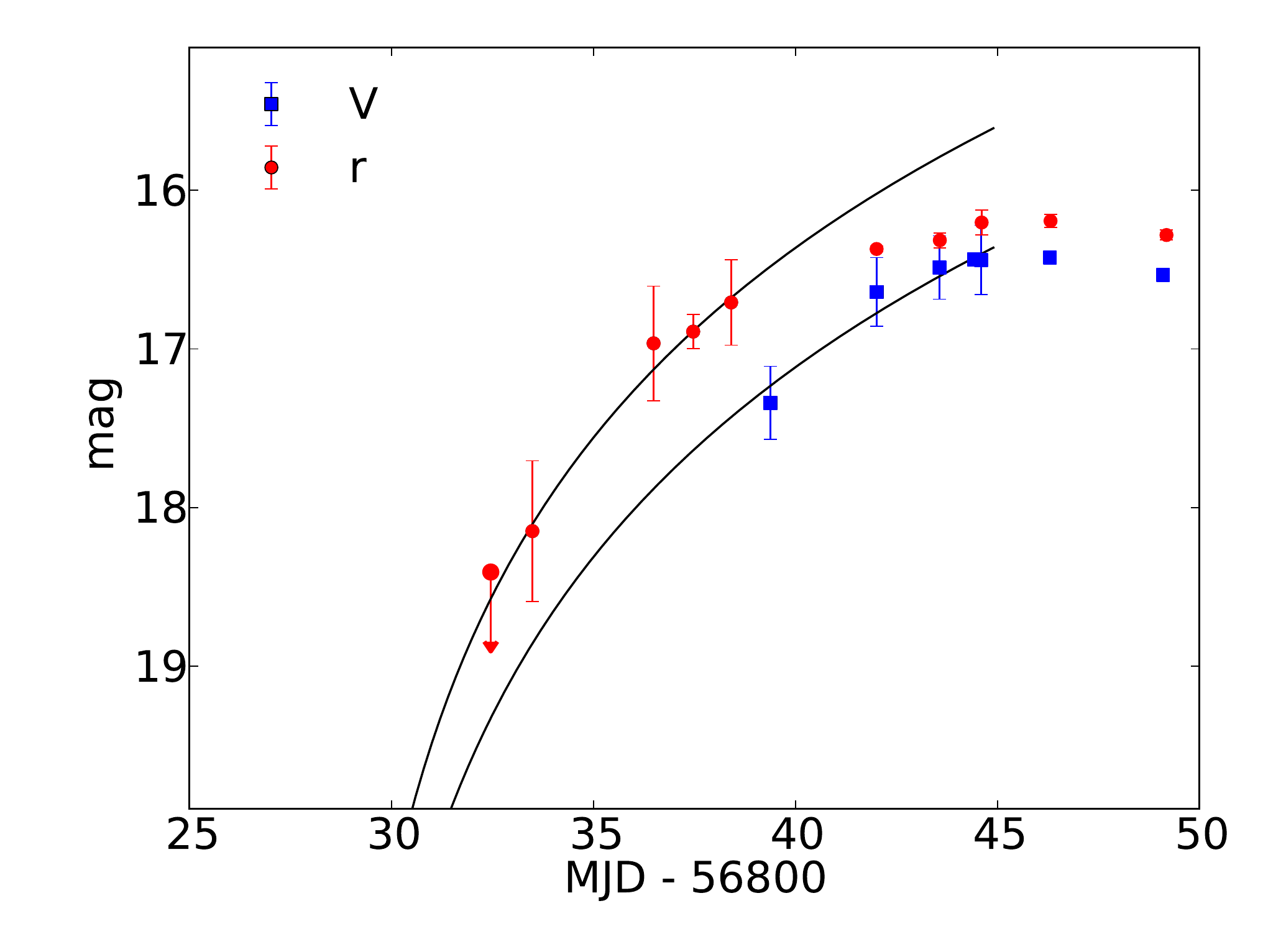}
\caption{Power-law fit to the 5 pre-maximum $r$- and $V$-band points using an ``expanding fireball'' model (index of the power law $n=2$). For comparison, a few post-maximum epochs are also shown, although they are not included in the fit. (A colour version of this figure is available in the online journal).} \label{fit_rise}
\end{figure}

The early detection of SN~2014ck and the analysis of LOSS/KAIT pre-discovery and discovery images gives a unique opportunity to obtain an accurate estimate of the rise time for a SN~Iax. 
In order to  constrain the explosion date of SN~2014ck, we fit the pre-maximum portion of the $r$- and $V$-band light curves (5 epochs for each band) with  an ``expanding fireball'' model, i.e. $f_{\rm model}(t) = \alpha(t - t_0)^n$ with $n=2$ \citep[][and references therein]{riess:1999}. The time of the first light ($t_0$) obtained from the fit (see Figure~\ref{fit_rise}) is ${\rm MJD} = 56828.2^{+2.7}_{-4.5}$ (2014 June 20.2 UT). With regards to the index of the power law $n$, \cite{firth:2015} presented an analysis of the early, rising light curves for a sample of 18 SNe~Ia: their data highlighted in many cases a departure of $n$ from the simple fireball model ($n=2$), with significant diversity from event to event (cf.\ their Table~4 and Figure~14) and a mean value of the distribution of $n = 2.44 \pm 0.13$. \cite{gane:2011}, using a sample of about sixty low-redshift LOSS SNe~Ia, found a best fit of $n = 2.20^{+0.27}_{-0.19}$, consistent ($1\sigma$) with the expanding fireball modelled by a parabola. In any case, these recent studies provide evidence for a range of $n$ for SNe~Ia events with the centre of the distribution slightly above 2 \citep[see also][]{piro:2014}. This deviation has implications for the distribution of $^{56}$Ni throughout the SN ejecta and so, in principle, the fit of the light curves should be $n$-free. Unfortunately, pre-maximum data for SN~2014ck are not enough for a convergent solution with a free $n$ parameter. To account for possible deviations from the fireball model, we fit a range of power laws with $2 \leq n \leq 2.5$ to the pre-maximum $r$- and $V$-band light curves independently. The reported uncertainty on $t_0$ is the standard deviation of the $t_0$ parameters for these fits.
We note that the spectral phases obtained by comparing the early spectra of SN~2014ck with similar-phase spectra of SN~2005hk are consistent with this estimate.

The rise-time 
to maximum is estimated to range from $\sim 17$~days in the $B$-band to $\sim 21$~days in the $i$-band. 
The $BVgri$-band rise time estimates are listed in Table~\ref{fit}. The associated errors are largely dominated by the error on $t_0$.  
From the bolometric luminosity (see Section~\ref{bolometric}) we infer a rise time of $t_{\rm rise} = 16.9^{+4.5}_{-2.7}$~days, in agreement with the $B$-band rise time. 
This is not surprising, as the $B$-band roughly traces the bolometric behaviour \citep{gane:2011}.  
Note the rise times for SNe~Iax range from SN~2008ha, at $\sim 10$~days, 
to SN~2008ge, which might be $> 20$~days \citep{foley:2013}.

\begin{table*}
\caption{Optical light curve parameters for SN 2014ck, with associated errors in parentheses.}\label{fit}
\begin{tabular}{cccccc}
\hline 
Filter & Peak & $m_{Peak}$ & $M_{Peak}$ & $\Delta m_{15}$ & $t_{\rm rise}$ \\
          &  MJD       & [mag]  &  [mag]  & [mag] & [days] \\
\hline 
$B$ & 56845.05 (0.50) & 16.87 (0.01) & $-$17.37 (0.15)&1.76 (0.15)&16.9$^{+4.5}_{-2.7}$\\   
$g$ & 56846.31 (0.15) & 16.65 (0.04) & $-$17.42 (0.15) &1.59 (0.10)&18.1$^{+4.5}_{-2.7}$\\   
$V$ & 56845.60 (0.10) & 16.41 (0.01) & $-$17.29 (0.15)&0.88 (0.05)&17.4$^{+4.5}_{-2.7}$\\
$r$  &  56846.62 (0.20) & 16.20 (0.03) & $-$17.29 (0.15)&0.58 (0.05)&18.4$^{+4.5}_{-2.7}$\\
$i$  & 56849.20 (0.60) & 16.08 (0.02) & $-$17.04 (0.15)& 0.39 (0.15)&21.0$^{+4.5}_{-2.7}$\\
\hline 
\end{tabular}
\end{table*}


\subsection{Bolometric light curve and explosion parameter estimates}\label{bolometric}

\begin{figure}
\includegraphics[scale=.44,angle=0]{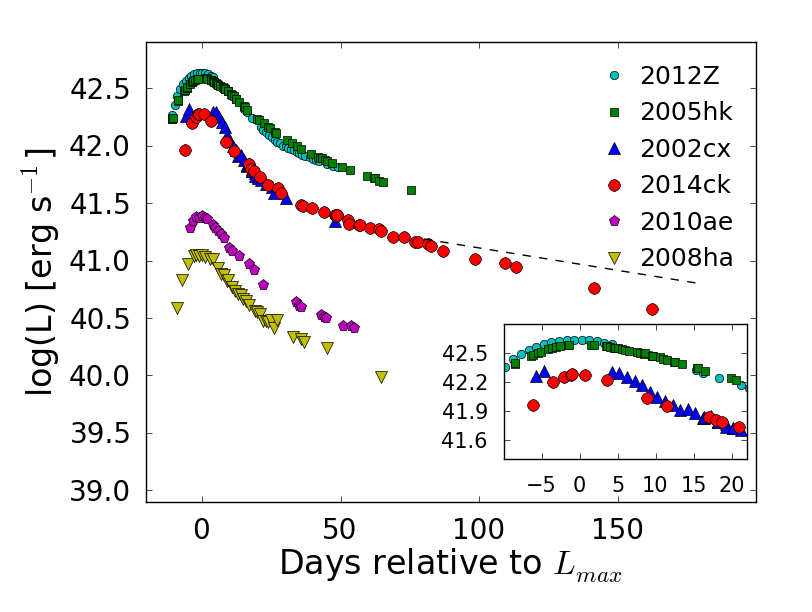}
\caption{OIR bolometric light curve of SN~2014ck, computed by integrating the fluxes from the $uBVgrizJHK$-bands. For comparison the OIR lightcurves are also shown for the Type~Iax SNe~2002cx \citep{li:2003,phillips:2007}, 2005hk \citep{phillips:2007,stritz:2015}, 2008ha \citep{foley:2009,valenti:2009,stritz:2014}, 2010ae \citep{stritz:2014} and 2012Z \citep{stritz:2015}. (A colour version of this figure is available in the online journal).} 
\label{bol}
\end{figure}

Using the multi-band photometry of SN~2014ck extending from  $u$- to $K$-band, we 
constructed the pseudo-bolometric optical-infrared (OIR) light curve shown in Figure~\ref{bol}. 
Unfortunately, no ultraviolet observations of SN~2014ck are available to compute a UVOIR bolometric light curve\footnote{The abbreviation UVOIR is used with different meanings in the literature. In this paper use it to mean the flux integrated from 1600 \AA\ ({\it Swift}/UVOT $uvw2$-band) to 25~\micron\ ($K$-band). If the integration starts from 3000 \AA\ (ground based $U$/$u$-band) we use the label OIR.} 

For each epoch and passband the observed magnitude was converted to flux at the effective wavelength.  
If observations were not available for a given filter on a particular night, the magnitude was estimated through interpolation between adjacent epochs or, if necessary, extrapolated assuming a constant colour from the closest available epoch. 
The fluxes were next corrected for reddening ($E(B-V)_{\rm tot} \approx 0.5 \pm 0.1$~mag), yielding  a full  spectral energy distribution (SED) at each epoch. 
The SEDs were then integrated using a trapezoidal integration technique, assuming zero flux at the integration boundaries (the edges of $u$ and $K$ bands). 
Finally, the fluxes at each epoch were converted to luminosities assuming our adopted distance to the host galaxy.

For comparison, the OIR pseudo-bolometric light curves of the  Type~Iax SNe~2005hk \citep[][adopting $E(B-V) = 0.11$~mag, $\mu = 33.46 \pm 0.27$~mag]{phillips:2007,stritz:2015}, 2012Z \citep[][$E(B-V) = 0.11$~mag, $\mu = 32.59 \pm 0.09$~mag]{stritz:2015}, 2010ae \citep[][$E(B-V) = 0.62$~mag, $\mu = 30.58 \pm 0.58$~mag]{stritz:2014} and 2008ha \citep[][$E(B-V) = 0.30$~mag, $\mu = 31.64 \pm 0.15$~mag]{valenti:2009,foley:2009,stritz:2014} were computed with the same prescription, using the optical and NIR photometry found in the literature. 

For SN~2002cx, only $BVRIr$ photometry is available \citep[][$E(B-V) = 0.034$~mag, $\mu = 35.09 \pm 0.32$~mag]{li:2003}, but given the striking photometric similarities between SNe~2002cx and 2014ck (see Figures~\ref{lc2} and \ref{col}), we assume the $u$- and NIR bands give the same contribution to the total flux (at least near maximum light) for both  objects. This contribution was estimated from 
the ratio in flux between the OIR 
and $BVri$-band bolometric light curves constructed for SN~2014ck, which is around 1.35 at maximum light
and decreases to 1.08 ten days after maximum, and applied to SN~2002cx.

\begin{figure*}
\includegraphics[scale=.6,angle=0]{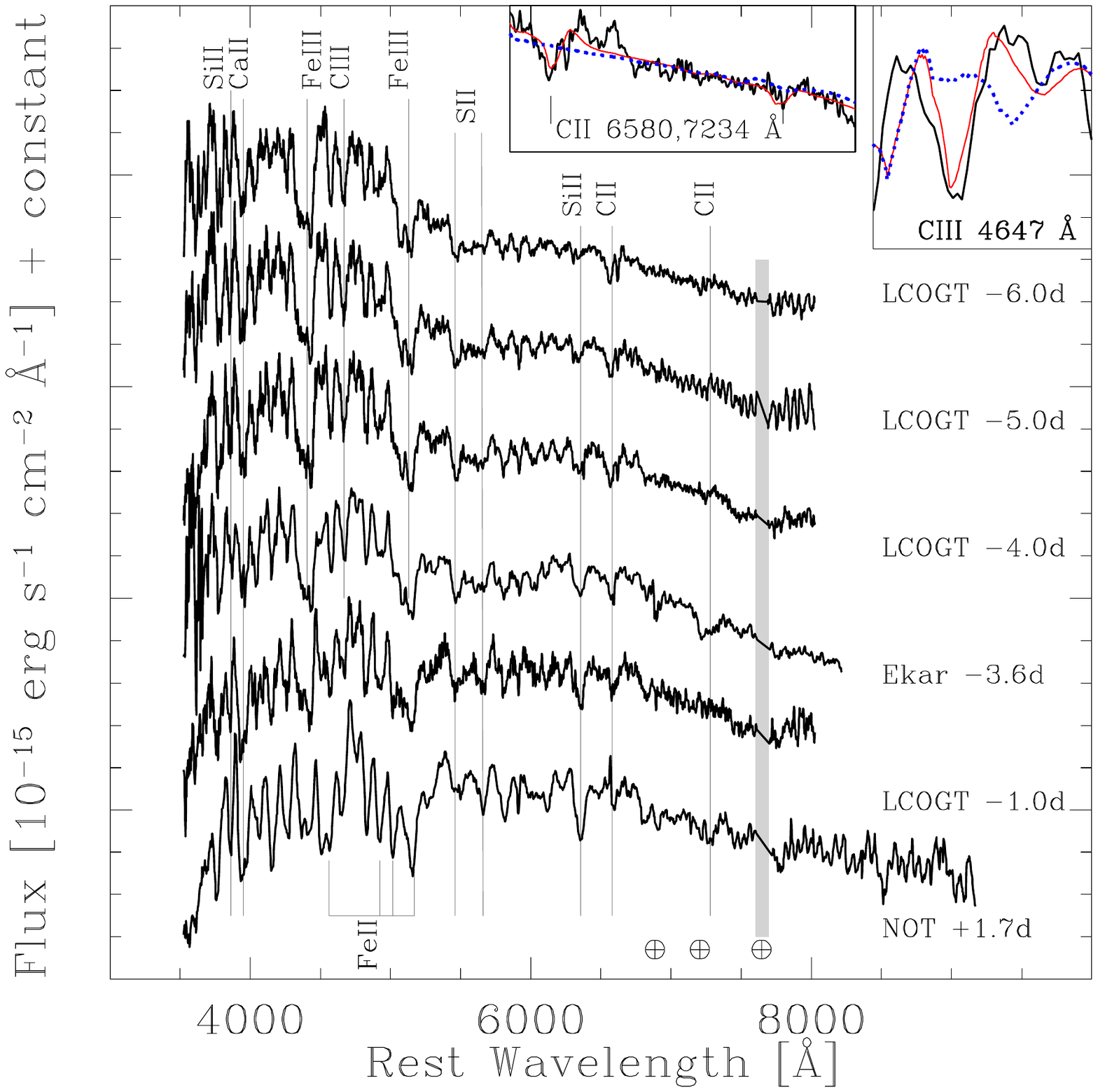}
\caption{Early spectral evolution and line identification. Phases relative to $V$-band maximum are reported. The insets on the top show regions of the $-4$~d spectrum centred on C~III $\lambda$4647 (right) and C~II $\lambda\lambda$6580, 7234 (left), with the synthetic spectra over-plotted, calculated with (solid red curve) and without (dotted blue curve) C~II and C~III ions (see text for details). Wavelength is in the rest frame, and the positions of major telluric absorption lines are marked with the $\oplus$ symbol (in particular the vertical grey band marks the strong O$_{2}$ $A$-band absorption at 7600 $\AA\/$). (A colour version of this figure is available in the online journal).} 
\label{spec_early}
\end{figure*}

Assuming that the light curve is powered by energy deposition from the  $^{56}{\rm Ni} \rightarrow {}^{56}{\rm Co} \rightarrow {}^{56}{\rm Fe}$ radioactive decay chain, the amount of $^{56}$Ni synthesised during the explosion can be estimated using Arnett's rule \citep{arnett:1982}. We applied the \cite{stritz:2005} analytic expression for deriving the energy input from the decay of $^{56}$Ni evaluated at the time of bolometric maximum (their Eq.~6). From the observed peak luminosity of SN~2014ck, $L_{\rm max}=1.91_{-0.26}^{+0.30} \times 10^{42}$~erg~s$^{-1}$, and rise time, $t_{\rm rise} = 16.9^{+4.5}_{-2.7}$ days (cf. Sections~\ref{photo_evol} and \ref{rise}), we obtain 
$M_{\rm Ni} \simeq 0.09^{+0.04}_{-0.03} M_\odot$. The uncertainty includes the error both in the determination of the rise time and in the adopted distance, which contribute $\sim 20\%$ and $\sim 16\%$, respectively, to the total error budget of the bolometric flux.

In principle, the contribution of UV light to the bolometric luminosity of SNe~Ia can be significant, particularly at the earliest epochs when the high temperature yields a large UV flux \citep{brown:2009,brown:2015}, affecting the calculated amount of $M_{\rm Ni}$. In the absence of UV data for SN~2014ck, it is interesting to note that SN~2005hk was already fading in the UV when {\it Swift}  observations began, nearly 10 days before the optical maximum \citep{phillips:2007,brown:2009}. Similarly, the UV light curves of SN~2012Z reach maximum well before the optical light curves \citep[][]{stritz:2015}. For both of them, the bolometric flux is 
dominated by the optical flux; the flux in the UV drops well below 10\% of the total flux before maximum. The same percentage was found for normal SNe~Ia \citep{suntzeff:1996,contardo:2000,brown:2009}.  
Thus, considering a maximum additional correction of $\sim 10\%$ for the contribution of the UV flux at $L_{\rm max}$, the $M_{\rm Ni}$ estimate for SN~2014ck increases to  
 $\simeq 0.10^{+0.04}_{-0.03} M_{\odot}$, but remains significantly lower than the typical values for normal SNe~Ia \citep[$\sim 0.2$ to $0.8M_\odot$, see][] {stritz06,hayden:2010}.
 
The rise time inferred for SN~2014ck and the extremely low expansion velocity of the ejecta ($v_{\rm ph} \simeq 3.0 \times 10^3$~km~s$^{-1}$, see Section~\ref{spec_evol}), suggest low ejecta mass ($M_{\rm ej}$) and kinetic energy ($E_{\rm k}$) compared to normal SNe~Ia and also to SN~2002cx \citep[for which $v_{\rm ph} \simeq 6.0 \times 10^3$~km~s$^{-1}$, see][]{li:2003}. Using  Arnett's equations \citep{arnett:1982} as per \cite{valenti:2008} -- a typo in their Eq.~2 was corrected by Wheeler et al. (2015) -- the OIR bolometric light curve is consistent with $M_{\rm ej} \sim 0.2$ to $0.5 M_\odot$, placing SN~2014ck close to the cluster made of Type~Iax SNe~2002cx, 2008A, 2005hk and 2009ku, just below the fast declining peculiar 1991bg-like SNe~Ia, in the $M_{\rm ej}$ vs.\ $M_{\rm Ni}$ plane plotted in Figure~15 of \cite{mccullyb:2014}.


With regard to the reliability of the $M_{\rm ej}$ estimate, it is well known that the opacities have a strong dependence on the temperature, and therefore that they vary with time \citep{hoeflich:1992}. Hotter, more luminous events should be characterised by higher opacities \citep{hoeflich:1996,nugent:1997,pinto:2001,maeda:2003,baron:2012}. 
We stress that molecules, such as CO, that are predicted to form efficiently in SNe~Iax \citep[and, in general, in sub-luminous SNe~Ia, see][]{hoeflich:1995}, do not provide significant opacity in the OIR spectral range. Hence, the above value of $M_{\rm ej}$ should be considered a lower limit, as discussed by \cite{stritz:2015} for SN~2012Z. 

\section{Spectral evolution}\label{spec_evol}

\subsection{Optical spectroscopy from $-6.0$ to $+110$  days}\label{spec_opt}

\begin{figure*}
\centering
\includegraphics[width=8.8cm]{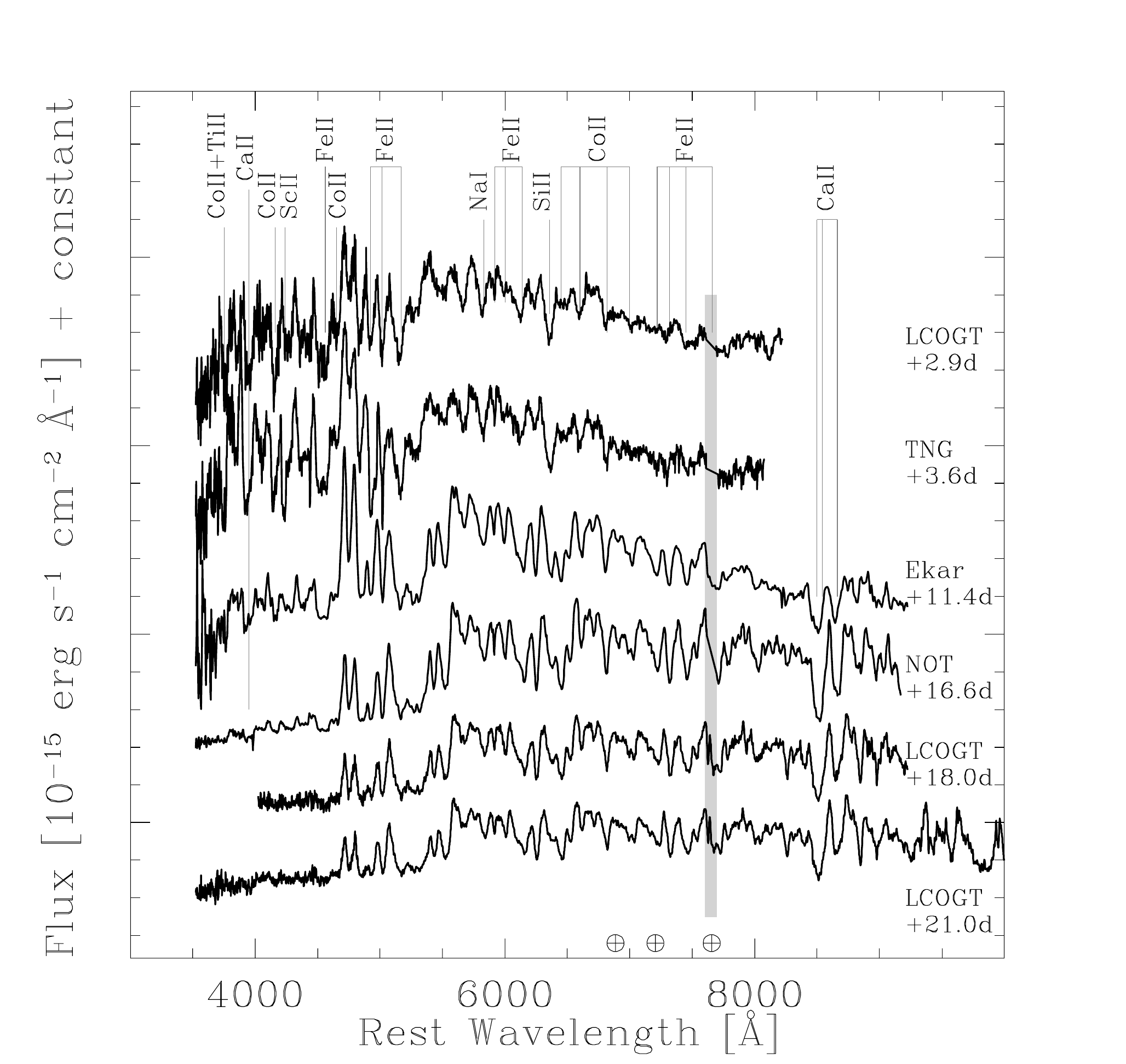}
\includegraphics[width=8.8cm]{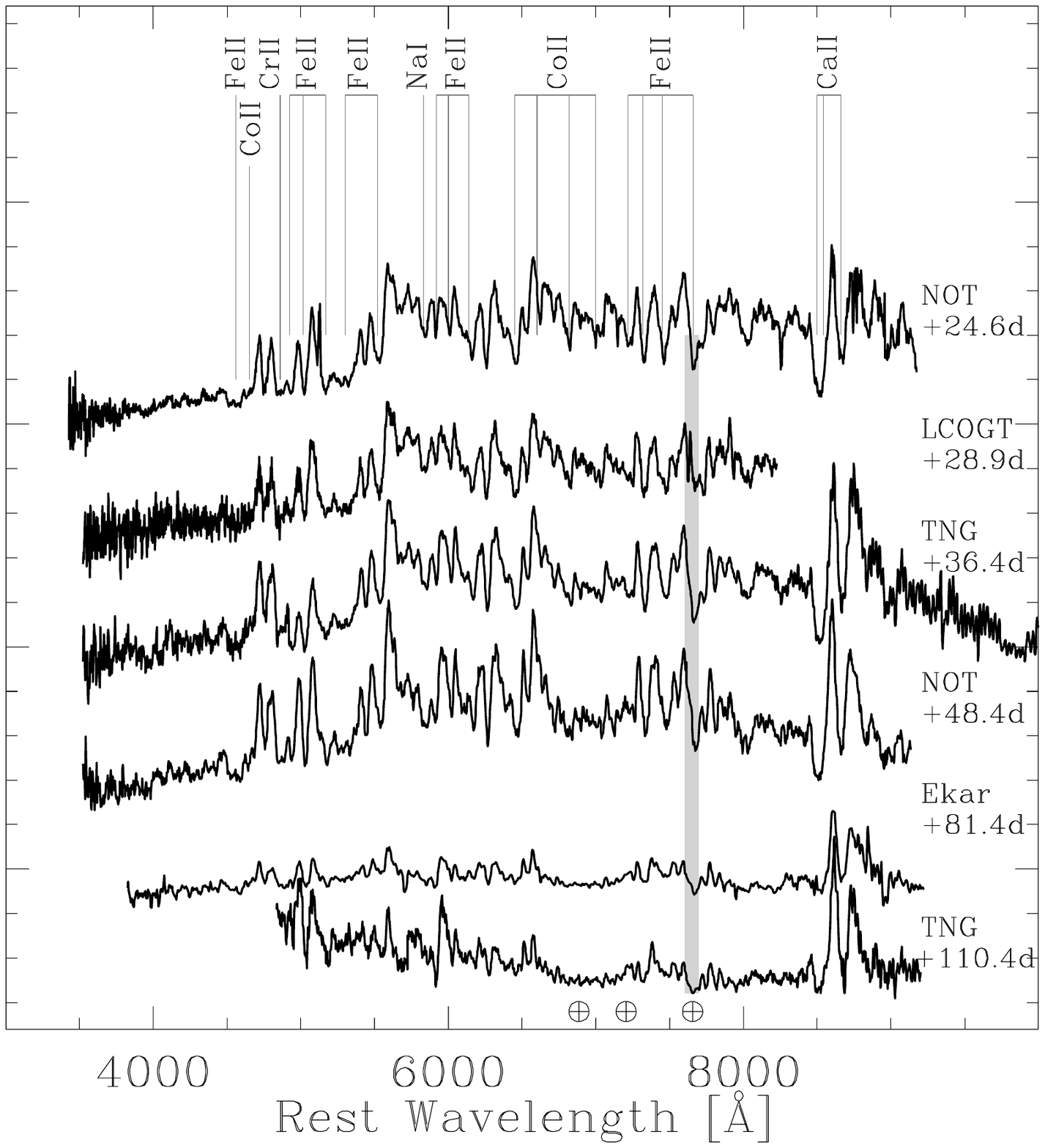}
\caption{Spectral evolution and line identifications. Phases are reported relative to $V$-band maximum. Left panel: spectra between  $+2.9$~d and $+21$~d. Right panel: spectra between  $+24.6$~d and $+110.4$~d. Wavelength is in the rest frame and the positions of major telluric absorption features are marked with the $\oplus$ symbol (in particular the vertical grey band marks the strong O$_{2}$ $A$-band absorption at 7600 $\AA\/$).  }
\label{fig_spectra1}
\end{figure*}

\begin{figure*}
\includegraphics[scale=.6,angle=0]{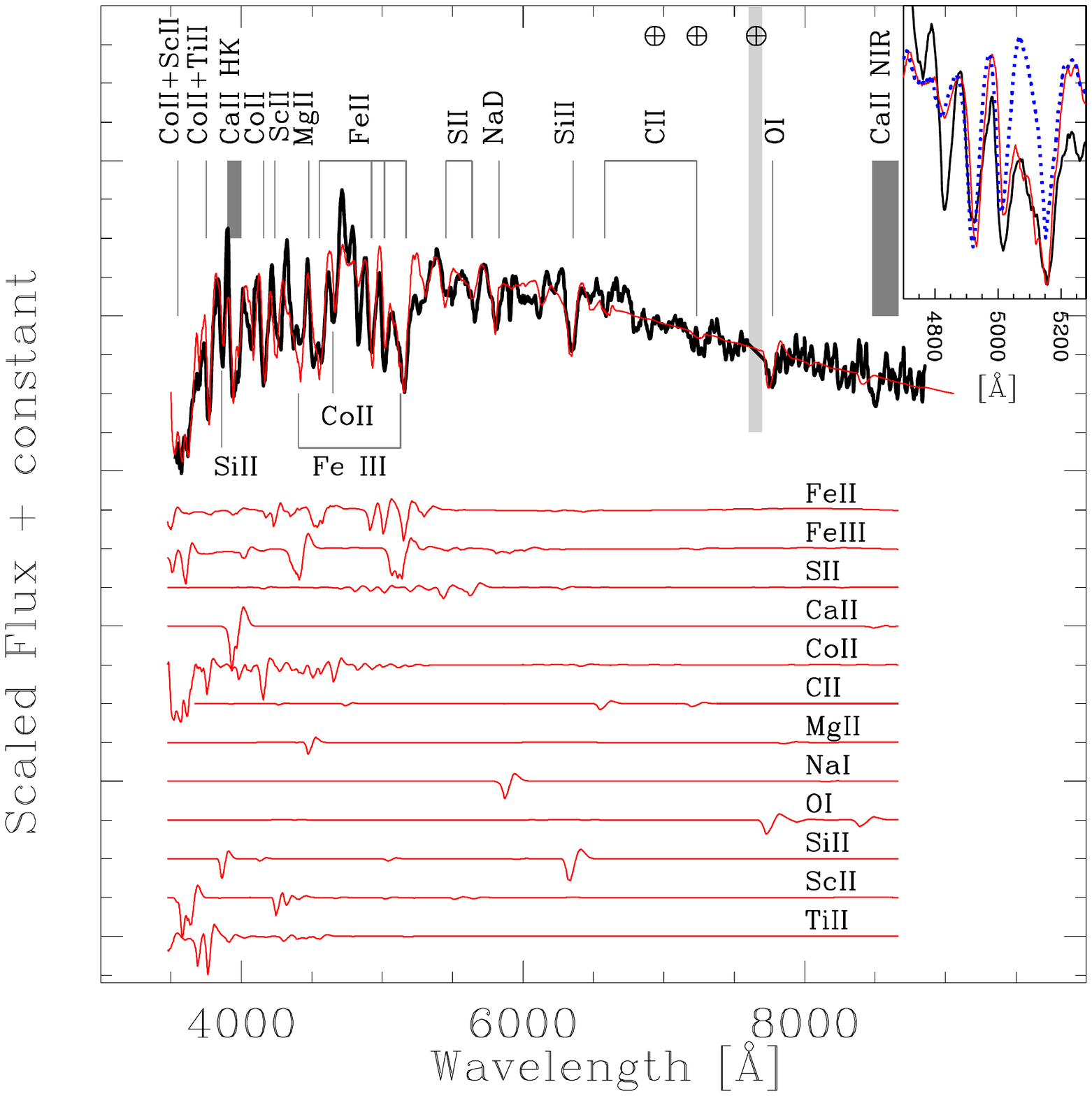}
\caption{Optical spectrum of SN~2014ck at $+1.7$~d (black) and our best-fit {\tt SYNOW} synthetic spectrum (red). The contribution of each ion is also shown. Major telluric features are indicated with the $\oplus$ symbol (in particular the vertical grey band marks the strong O$_{2}$ $A$-band absorption at 7600 $\AA\/$). The insets on the right of the plot show the regions around Fe~III $\lambda$4404 (bottom) and Fe~III $\lambda$5129 (top) with an synthetic spectrum calculated without Fe~III over-plotted (dotted blue). (A colour version of this figure is available in the online journal).} 
\label{synow}
\end{figure*}

\begin{figure}
\includegraphics[scale=.43,angle=0]{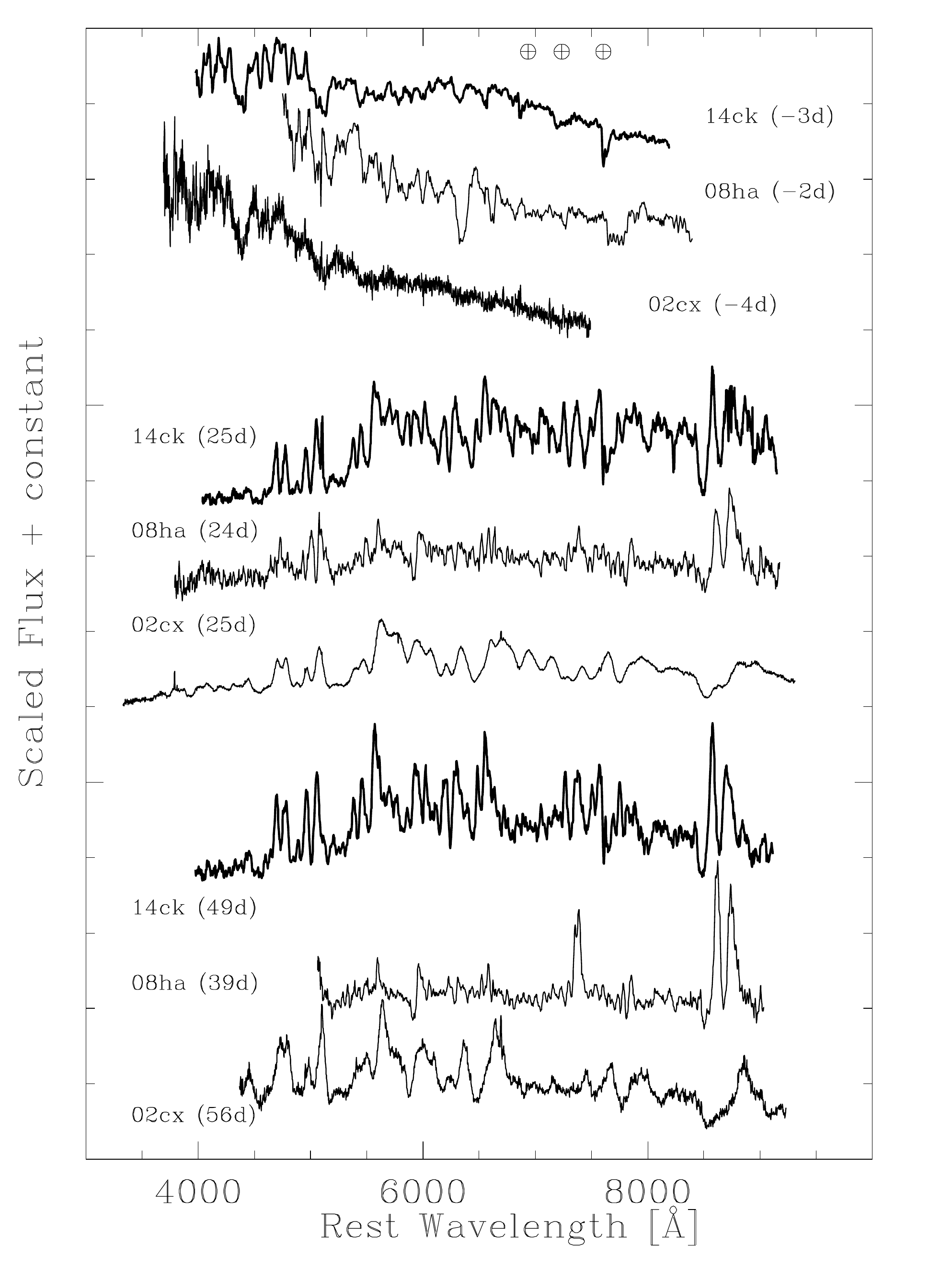}
\caption{Comparison of the rest-frame spectra of SN~2014ck at phases $-3.5$~d, $+25$~d and $+48$~d with those of SNe~2002cx \citep{li:2003,phillips:2007} and 2008ha \citep{valenti:2009,foley:2009} at similar phases.} 
\label{spec_conf}
\end{figure}

The spectral evolution of SN~2014ck at optical wavelengths is shown in Figures~\ref{spec_early}, \ref{fig_spectra1} and \ref{fig_spectra4}. There is no sign of helium or hydrogen features. The pre-maximum spectra plotted in Figure~\ref{spec_early} exhibit a blue continuum with a weak, narrow Si~II~$\lambda$6355 absorption line -- the hallmark of SN~Ia -- as well as Fe~III~$\lambda\lambda$4404, 5129 and a relatively strong feature at 4670~\AA, tentatively identified as C~III~$\lambda$4647. 
\citep[There is some indication that the early spectra of SNe~2005hk and 2012Z contain C~III; see][]{chornock:2006,foley:2013}. 
At the blue end of the spectra, Ca~II~H\&K and Fe~II~$\lambda$4555 are also identified, while redward of 5000~\AA,  features associated with S~II~$\lambda\lambda$5454, 5640 and C~II~$\lambda\lambda$6580, 7234 are detected. C~II absorption lines have been reported in SN~2008ha \citep{foley:2010b} and possibly identified in SNe~2002cx \citep{parrent:2011}, 2005hk \citep{chornock:2006}, 2007qd \citep{mcclelland:2010}, 2012Z \citep{stritz:2015,yamanaka:2015} and several other SNe~Iax \citep[][their Figure~23]{foley:2013}. 

To verify the consistency of line identification and photospheric parameters, we make use of the parametrised synthetic-spectrum code {\tt SYNOW} \citep{fisher:1997,branch:2002,branch:2003,branch:2004}. In short, {\tt SYNOW} generates a synthetic spectrum, starting from an adopted blackbody continuum temperature ($T_{\rm bb}$), photospheric expansion velocity ($v_{\rm ph}$) and, for each selected ion, a few specific parameters \citep[i.e. the optical depth of the reference line; the excitation temperature; the minimum, maximum and optical-depth e-folding velocities; see][]{branch:2002,branch:2004}. 
With {\tt SYNOW}, synthetic spectra were computed to match the observations at different epochs using ions believed to be present in SN~2014ck, following \cite{branch:2004,jha:2006,chornock:2006,sahu:2008}. In particular we included iron-peak elements, intermediate-mass elements and unburned carbon. 

Examples of {\tt SYNOW} spectra are shown in the insets of Figure~\ref{spec_early} (phase $-4$~d) and in Figure~\ref{synow} (phase $+1.7$~d). 
For the pre-maximum spectra, $T_{\rm bb} \approx 7000$~K and $v_{\rm ph} \approx 3500$~km~s$^{-1}$ (see Section~\ref{bb}) were adopted.
The parameters for the fit to the $+1.7$~d spectrum are $T_{\rm bb} = 5600$~K and  $v_{\rm ph} = 3000$~km~s$^{-1}$. 
We included the set of ions and input parameters used by \cite{branch:2004} for the analysis of the early spectra of 
SN~2002cx, i.e. Fe~II, Fe~III, Co~II, Cr~II, Ca~II, Na~I, Si~II, Si~III, S~II and Ti~II (see their Tables~1 and 2).  
C~III and C~II ions were added in our {\tt SYNOW} spectral model to obtain reasonable fits to absorption features at $\sim 4650$~\AA\ and $\sim 6580$ and 7230~\AA, respectively. This is shown in the insets in the top panel of Figure~\ref{spec_early}, where the synthetic models obtained with and without C~III and C~II ions are compared to the observed $-4$~d spectrum. 
Spectra obtained near maximum light contain Fe~III features (see the insets around Fe~III $\lambda\lambda$4404, 5129 in Figure~\ref{synow}), while 
soon after maximum Fe~III lines have vanished and strong Fe~II lines have developed \citep[as already noted for SN~2002cx by][]{branch:2004}
  
We tried also to include Sc~II, Ni~I and Ni~II, instead of Fe~II and Co~II, which might contribute with features blueward of 4000~\AA\ (especially Sc~II). 
However, lines of Fe~II produce most of the observed features, and line blanketing by Co~II lines is needed to get a reasonable fit in the blue. 
Na~I, Ca~II, Mg~II and O~I produce just one feature each (see Figure~\ref{synow}). 

Carbon is very likely overabundant in the outer layer of SN~2014ck, since in the very early spectra C~II and C~III are the strongest lines, along with Fe~III. 
The detection of unburned (C+O) material in the ejecta and, specifically, the spectroscopic signatures and velocity structure of C, is of great importance for constraining our understanding of the explosion mechanisms. In particular, it might be related to the mechanism by which the explosive flame propagates throughout the WD star \citep{parrent:2011,folatelli:2012} and/or the type of progenitor system \citep[C/O WD or O/Ne/Mg WD, see ][]{nomoto:2013}. Actually, the hallmark of pristine material from C/O WD progenitor star is the presence of carbon, as oxygen is also a product of carbon burning.
For SN~2014ck, the measured pseudo-EW \citep[see][]{folatelli:2012} of 
C~II~$\lambda$6580 is $\sim 10$~\AA\ at phase $-6$~d, which decreases to $\sim 4$~\AA\ two days after $V$-band maximum. 
For comparison, for normal SNe~Ia, \cite{folatelli:2012} find a C~II~$\lambda$6580 pseudo-EW of about 4~\AA\ five days before maximum and $\sim 1$~\AA\ at maximum. 
This supports the findings from the analysis of SNe~2005cc and 2008ha, presented by \cite{foley:2013}, where the C~II~$\lambda$6580 signature is quite strong \citep[see also][]{parrent:2011}: a large fraction of unburned material is present in the ejecta of at least some SNe~Iax, and almost every SN~Iax with a spectrum before or around maximum light has some indication of carbon absorption. 

Taking the Si~II $\lambda$6355 absorption line as indicator of the photospheric velocity at early epochs \citep{patat:1996}, 
the ratio between the Doppler velocity of C~II~$\lambda$6580 and Si~II~$\lambda$6355 (see Section~\ref{bb}, Table~\ref{tbb_vel}) at maximum is $\sim 0.95$. It was $\sim 0.89$ for SN~2012Z \citep[][their Figure~9]{stritz:2015} and around 0.6 for SN~2008ha \citep{parrent:2011}. This ratio is generally slightly above unity among SNe~Ia \citep{parrent:2011,folatelli:2012}, indicating either a layered distribution of carbon-rich material or multiple clumps having similar filling factors.  On the other hand, a Doppler velocity of C~II significantly below that of the photospheric velocity may indicate ejecta asymmetries, as might be the case for SN~2008ha.

The post-maximum spectra show the emergence of several Fe~II (and even Co~II) lines becoming dominant over a two-week period. Ni~I and Ni~II might also contribute blueward of 4000~\AA, likely blended with numerous Fe~II and Co~II lines. 
The Si~II~$\lambda$6355 feature is clearly visible until 15~days after maximum brightness, as in SN~2002cx \citep{li:2003,branch:2004}. 
On the other hand, in the case of SN~2008ha and other faint SN~Iax, this feature is only visible near maximum light \citep{valenti:2009,foley:2009,foley:2010b,stritz:2014}. Carbon features are clearly detected before maximum. 
From $+24.6$~d to $+110.4$~d the spectra are dominated by Fe~II and Co~II lines, as well as by the progressive emergence of the Ca~II~NIR triplet.

In Figure~\ref{spec_conf} the spectra of SN~2014ck are compared to those of SNe~2002cx and 2008ha at similar phases. 
Notably, the pre-maximum spectrum of SN~2014ck resembles SN~2008ha (rather than SN~2002cx), with the exception of the Si~II~$\lambda$6355 absorption line which is clearly stronger in SN~2008ha.  
Twenty-five days after maximum brightness, the Ca~II NIR triplet in SN~2014ck is as strong as in SN~2008ha \citep{valenti:2009}, while this feature is much weaker in SN~2002cx. Around fifty days after maximum, [Ca~II]~$\lambda\lambda$7291, 7324 emission lines begin to emerge. At similar phases, these forbidden lines are stronger in SN~2008ha and extremely weak in SN~2002cx. 
Overall, the spectra of SN~2014ck show a strong similarity to SN~2008ha and clear differences from SN~2002cx, particularly due to the smaller expansion velocities, but possibly also due to different ejecta composition and opacity.

The very low expansion velocity of SN~2014ck may enhance the visibility of Sc~II, tentatively identified in the narrow-line SNe~2007qd \citep{mcclelland:2010} and 2008ha \citep{valenti:2009,foley:2009}. 

\subsubsection{Expansion velocities of the ejecta}\label{bb}

\begin{table*}
\caption {Blackbody temperatures (Kelvin) and expansion velocities of the ejecta (km~s$^{-1}$) at the absorption feature minimum for various ions in SN~2014ck. Estimated uncertainties are in parentheses. Phase is from the adopted epoch of the $V$-band maximum, ${\rm MJD} = 56845.6 \pm 0.1$.}
\tiny
\label{tbb_vel}
\begin{tabular}{ccccccccccccccc}
\hline 
Phase	&$T_{\rm bb}$& Si II       &  Ca II& Ca II       & C II          &C II         &S II& O I    &Fe III             &Fe II  & Fe II                                & Co II          & Co II          & Co II      \\
(d)     &         (K) &$\lambda6355$&  H\&K &$\lambda8498$& $ \lambda6580$&$\lambda7234$&$\lambda\lambda5454,5640$ &$\lambda7774$& $\lambda5129$ &$\lambda6149$& $ \lambda6247$ & $ \lambda15759$& $\lambda16064$ &
$\lambda16361$ \\
\hline                                   
$-$6.0 &   8140(100)&  3445(50)& 4110(200)&     3970(200)&  3390(50) &  3460(100)&3180(100)& 3826(50)&3200(100)   &5300(200)    &   4740(200) &     $-$     &  $-$     &$-$\\
$-$5.0 &   7340(100)&  3339(50)& 3800(100)&     3760(200)&  3030(50) &  $-$      &2797(100)& 3200(50)&3000(100)   &5200(200)    &   4700(200) &     $-$     &  $-$     &$-$  \\
$-$4.0 &   6720(100)&  2950(50)& 3540(100)&     $-$      &  2935(100)&  3130(100)&2395(100)& 3020(50)&2820(100)   &4690(100)    &   4620(100) &      $-$    &  $-$     &  $-$  \\
$-$3.6 &   6900(100)&  2890(50)& 3450(100)&     3400(200)&  2800(50) &  2960(100)&  $-$  &    2962(50)&$-$        &4590(100)    &   4500(100) &    $-$      &   $-$    &  $-$    \\
$-$3.0 &   6300(200)&  2611(50)& 3305(200)&     $-$    &    2800(50) &  2960(100)&   $-$ &    2866(50)&  $-$      &4490(100)    &   4200(100) &     $-$     &$-$       &   $-$    \\
$-$1.0 &   5800(200)&  $-$     & $-$      &     $-$    &      $-$    &    $-$    &  $-$  &      $-$   &  $-$      &$-$          &  $-$        &      $-$    &  $-$     &$-$\\
$-$1.0 &     $-$    &      $-$ &$-$       &     $-$    &       $-$   &     $-$   &   $-$ &       $-$  &   $-$     &$-$          &   $-$       & 2700(200)  &2360(200)&2450(200)\\
$-$0.1    &     $-$    & 2610(100)& 3140(200)&     3100(200)&     $-$   &     $-$   &   $-$ &    2750(50)&  $-$      &4180(100)    &   3800(100) & 2656(200)  &2262(200)&2339(200)\\
1.2
7&   5630(200)& 2560(100)&     $-$  &     $-$  &         $-$   &     $-$   &   $-$ &    $-$    &   $-$     &4008(100)    &   3350(100) &    $-$      &  $-$     &$-$\\
2.9    &   5360(200)& 2430(100)&$-$       &     $-$  &         $-$   &     $-$   &   $-$ &    2600(50)&      $-$  &3800(100)    &   3160(100) &  $-$        &   $-$    &  $-$     \\
3.6    &   5360(200)& $-$      &   $-$    &     $-$  &         $-$   &     $-$   &   $-$ &      $-$   &  $-$      &     $-$     &    $-$      &  $-$       &   $-$    &$-$\\
3.9    &    $-$     & 2250(100)& 2800(200)&     2700(200)&     $-$   &     $-$   &   $-$ &       $-$  &   $-$     &3311(100)    &   3201(100) &  2613(200)  &2243(200)&2357(200)\\
5.0    &   5050(300)& $-$      &$-$       &      $-$    &   $-$      &  $-$      &$-$    &    2300(50)&      $-$  &2700(100)    &   2707(100) &   $-$        &$-$&    $-$   \\
11.4   &   4800(300)&$-$       &$-$       &      $-$    &   $-$      &  $-$      &$-$    &    1900(50)&      $-$  &2404(100)    &  2418 (100) &   $-$        &    $-$   &    $-$   \\
16.6   &   4500(300)&  $-$     &$-$       &      $-$    &   $-$      &  $-$      &$-$    &      $-$   &   $-$     &2224(100)    &   2313(100) &  $-$        &   $-$    &   $-$    \\
18.0   &   4100(300)&  $-$     &       $-$&      $-$    &   $-$      &  $-$      &$-$    &      $-$   &    $-$    &  $-$        &    $-$      &      $-$   &$-$&$-$\\
19.0   &$-$         &  $-$     &     $-$  &      $-$    &   $-$      &  $-$      &$-$    &      $-$   & $-$       &2137(100)    &   2180(100) & 1955(50)   &1965(50)&1938(50)\\
20.0   &   4110(300)&$-$       &$-$       &      $-$    &   $-$      &  $-$      &$-$    &       $-$  &  $-$      &2107(100)    &   2246(100) &  $-$        &  $-$     &$-$\\
21.0   &   4050(300)&     $-$  &$-$       &      $-$    &   $-$      &  $-$      &$-$    &     $-$    &   $-$     &     $-$     &    $-$      &  $-$        &$-$&$-$\\
23.7   &     $-$    &$-$       &$-$       &      $-$    &   $-$      &  $-$      &$-$    &      $-$   &$-$        &2079(100)    &   2146(100) & 1776(50)   &1835(50)&1865(50)\\
24.6   &   4070(300)& $-$      &   $-$    &      $-$    &   $-$      &  $-$      &$-$    &     $-$    &  $-$      &2011(100)    &   2132(100) & $-$        &$-$&  $-$     \\
28.9   &   4130(300)&$-$       &$-$       &      $-$    &   $-$      &  $-$      &$-$    &      $-$   &   $-$     &     $-$     &    $-$      & $-$       &$-$&  $-$     \\
30.8   &$-$         &$-$       &$-$       &      $-$    &   $-$      &  $-$      &$-$    &     $-$    &   $-$     &1740(100)    &   1769(100) & 1634(50)   &1742(50)&1683(50)\\
36.4   &   4110(300)&  $-$     &     $-$  &      $-$    &   $-$      &  $-$      &$-$    &      $-$   &   $-$     &1919(100)    &   1950(100) &$-$        &  $-$     &   $-$     \\
39.0   &    $-$     &$-$       &$-$       &      $-$    &   $-$      &  $-$      &$-$    &     $-$    &     $-$   &     $-$     &    $-$      &   $-$     & $-$      &  $-$      \\
46.8   &$-$         &     $-$  &$-$       &      $-$    &   $-$      &  $-$      &$-$    &      $-$   &      $-$  &1613(100)    &   1680(100) &     $-$     &$-$&    $-$    \\
48.4   &   4120(300)&  $-$     &$-$       &      $-$    &   $-$      &  $-$      &$-$    &    $-$     &       $-$ &1580(100)    &   1511(100) &       $-$    &$-$&$-$\\ 
\hline 
\end{tabular}
\end{table*}
   
\begin{figure}
\includegraphics[scale=.43,angle=0]{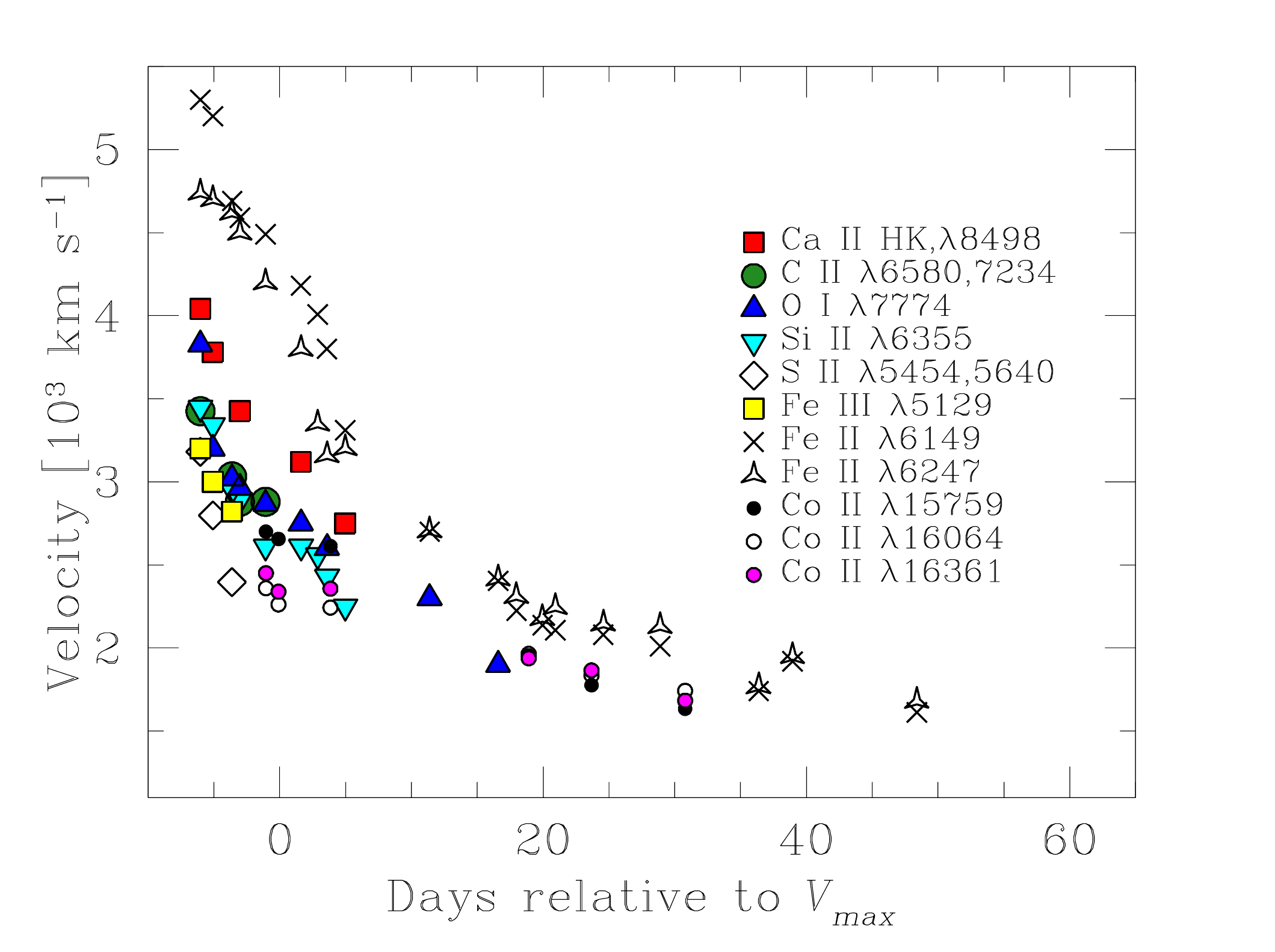}
\caption{Velocity evolution of the absorption minima of a selection of spectral lines with minimal line blending in the spectra of SN~2014ck. The typical formal error of the velocities is $\sim 200$~km~s$^{-1}$. The blending of lines can produce a systematic shift of features and increase the uncertainty at least to $\sim 1000$~km~s$^{-1}$.} 
\label{vel_lines}
\end{figure}

One of the main properties of SNe~Iax is the low expansion velocity of the ejecta, which suggests low explosion energies compared to normal SNe~Ia (see Section ~\ref{bolometric}). 
The ejecta velocity of SN~2014ck was estimated from the location of the absorption minima of spectral lines with little line blending, based on line identifications from {\tt SYNOW}.

The results are listed in Table~\ref{tbb_vel} and plotted in Figure~\ref{vel_lines}. Early spectra can be used to probe the velocity distribution of various elements, i.e. unburned (C+O) material, intermediate mass elements (IMEs) and completely burned elements close to nuclear statistical equilibrium (NSE), namely iron, cobalt and nickel. In principle, the velocity evolution can provide solid constraints on the explosion physics, as a layered structure might be revealed \citep[a signature of detonations,][]{stritz:2014,stritz:2015} unless extensive mixing has destroyed the original stratification \citep[a signature of deflagrations,][]{phillips:2007}. However, for SNe~Iax there is severe blending of lines over the full optical and NIR spectral range that prevent secure line identifications and plague our velocity estimates \citep{szalai:2015}. 

Before maximum light, expansion velocities are measured from the minima of Ca~II H\&K and $\lambda$8498; C~II $\lambda\lambda$6580, 7234; S~II $\lambda\lambda$5454, 5640; and Si~II $\lambda$6355 absorption features and are found to lie between 2800 and 4100~km~s$^{-1}$. Mg~II $\lambda$4481 soon becomes blended with Fe~III and is not easily distinguished. On the contrary, the O~I line at $\lambda$7774 is clearly detected redward of the telluric $A$-band at 7590-7650 \AA\ and the calculated velocities are similar to the ones inferred for Si~II. At these early phases, iron features Fe~II~$\lambda$6149, 6247 and 4555 exhibit consistent line velocities that are $\sim 1000$~km~s$^{-1}$ higher than those of IMEs and Fe~III~$\lambda5129$. However, we note that complex blending with emerging Co~II and Ti~II features could change the position of Fe~III~$\lambda5129$ (and also Fe~III~$\lambda4404$) absorption minima. 

Indeed, around maximum light, the blending with emerging Fe~II, Ti~II and Co~II lines might broaden the observed profile and shift the middles of several lines. In particular, even during pre-maximum phases, the feature around 6300 \AA, mainly attributed to Si~II $\lambda$6355, could be a blend with Fe~II $\lambda6456$ \citep[or a more complex blending either with Fe~II plus Co~II or S~II, see][their Figures~11 and 12]{szalai:2015}. 
After maximum, the ``iron curtain'' prevents the secure identification of unburned (C+O) material or IMEs, forming at similar or higher velocities \citep{branch:2004}, and the velocity measurements might be ill constrained. 
Close to maximum, the absorption-minimum velocities of the Co~II NIR lines at 1.5759, 1.6064 and 1.6361~\micron\ are in good agreement either with Si~II~$\lambda$6355 or S~II~$\lambda\lambda$5454, 5640, and are $\sim 2000$~km~s$^{-1}$ lower than for Fe~II~$\lambda6149, 6247$. 

About twenty days after maximum, the velocities of the Co~II NIR features are systematically 
$\sim 300$~km~s$^{-1}$ lower than those inferred from optical Fe~II lines at the same phase. A similar trend was noted by \cite{stritz:2014} for SN~2010ae. Hereafter the line velocities evolve rather slowly. 

Overall, the velocity structure of SN 2014ck indicates outer layers rich in iron-group ions, while C+O elements, Si~II and Ca~II, identified at lower velocities, seem to be well mixed. In principle, they should be present even at higher velocities (i.e. in the outer layers) if earlier spectra were available for an in-depth analysis. Consequently, we cannot exclude either a mixed or a layered structure for SN~2014ck and, in turn, it is not easy to discriminate between the different explosion mechanisms (see Section~\ref{discussion}).

In Table~\ref{tbb_vel} we also list our estimates of the photospheric temperature of SN~2014ck as derived from a blackbody fit to the spectral continuum (the spectra were corrected for the redshift and extinction). 
At phases beyond $+50$~d, emerging emission lines and line blanketing drive a flux deficit at the shorter wavelengths, and the fit becomes difficult. 
The errors are estimated from the dispersion of measurements obtained with different choices for the spectral fitting regions. 
The early photospheric temperature of SN~2014ck is above 8000~K, but it decreases quickly to $\sim 5600$~K at maximum light and flattens to about 4000~K afterwards.

\subsection{The late-time spectrum at $+166.3$~d}

\begin{figure*}
\includegraphics[scale=.8,angle=0]{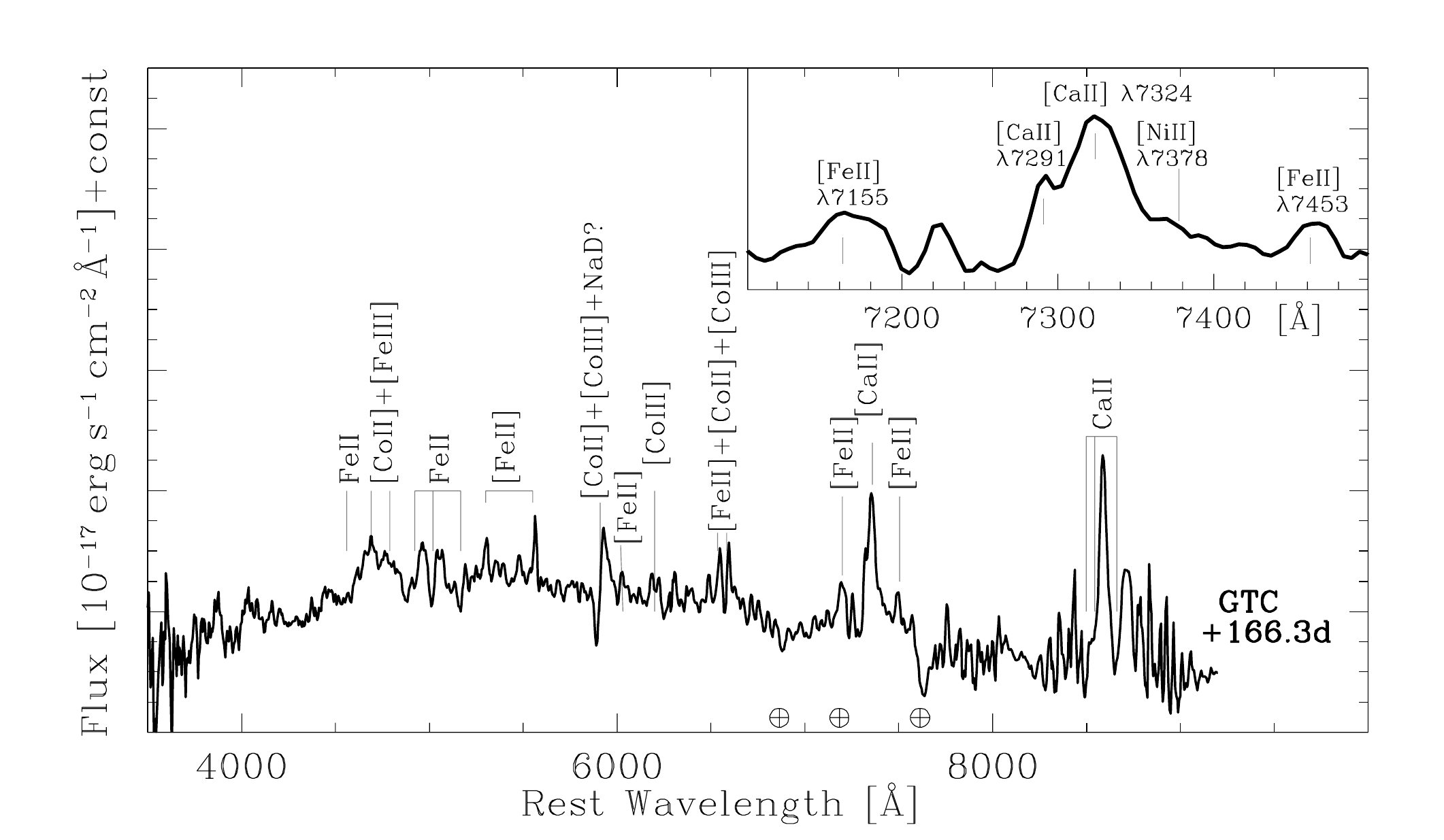}
\caption{Late-phase spectrum of SN~2014ck with line identification based on Li et al. (2003), Jha et al. (2006) and Sahu et al. (2008). The inset shows the region around $\sim 7300$~\AA, where forbidden [Ca II]~$\lambda\lambda$7291, 7324 and [Fe~II]~$\lambda\lambda$7155, 7453 features are clearly identified. [Ni~II]~$\lambda$7378 might be present in the red wing of [Ca II]~$\lambda$7324.} 
\label{fig_spectra4}
\end{figure*}

A late-phase spectrum of SN~2014ck was obtained at $+166.3$~d and is plotted in Figure~\ref{fig_spectra4}. 
As already remarked for other SNe~Iax \citep{foley:2013}, the late-phase spectrum does not appear to be truly nebular despite no clear indication of a continuum or absorption lines.
At these late epochs, SN~2014ck shows narrow permitted Fe~II lines superimposed on a pseudo-continuum and several forbidden lines associated with of Fe, Co and Ca. 
The dominant feature is the Ca~II~NIR triplet, but also comparable in strength is the forbidden [Ca~II]~$\lambda\lambda$7292, 7324 doublet (see Figure~\ref{fig_spectra4}). 
Both permitted and forbidden calcium lines are significantly more prominent in SN~2014ck than in SN~2002cx, and are comparable to those of the fainter SNe~2008ha and 2010ae \citep{valenti:2009,foley:2009,stritz:2014}.  
The [Fe~II] $\lambda$7155 line is the strongest iron feature in the late-time spectrum. A relatively broad hump at 4700 \AA\ is identified as [Fe~III] and [Co~II].

A number of features blueward of 5800~\AA\ are likely a blend of permitted Fe~II lines. 
The same lines are present at earlier epochs but at higher velocities. This identification was suggested by \cite{jha:2006} for SN~2002cx. 
As a test, we attempted to fit the late time spectrum with a {\tt SYNOW} model, including Fe~II, Ca~II, Na~I and O~I ions \citep[see][their Table~2 and their Figures~3 and 4]{jha:2006}. 
Although the observed spectrum is not fully reproduced by the photospheric model, the synthetic spectrum provides a good match to many of the
absorption features blueward of 5800~\AA, and possibly the P-Cygni profiles of Na~I~D at $\lambda\lambda$5890, 5896. An alternative identification of this feature could be [Co~III]~$\lambda$5888 \citep{dessart:2014}, also suggested for SN~2012Z by \cite{stritz:2015}.
This last interpretation is supported by the unambiguous presence of other [Co~III] lines in the 6000~\AA\ region.  

We conclude that the late-time spectrum of SN 2014ck is a combination of P-Cygni profiles of recombination lines and emission lines of forbidden Fe, Ca and Co features, but no [O~I]~$\lambda\lambda$6300, 6364 emission. This feature is typically not present in late spectra of SNe~Ia \citep{blondin:2012}.
In order to get [O~I] emission, we need a significant amount of O in a region where $\gamma$-rays are being absorbed, and the O-emitting region cannot be too contaminated by Ca. In fact, the [Ca~II]~$\lambda\lambda$7291, 7324 feature can limit the strength of [O~I]~$\lambda\lambda$6300, 6364  emission from a region in which both these ions co-exist \citep{fransson:1989,dessart:2015}. The emission of [O~I]~$\lambda\lambda$6300, 6364 is absent from relatively late spectra of SNe~2014ck and 2008ha \citep{foley:2009}, while in both cases O~I~$\lambda$7774 absorption is identified in photospheric phase spectra.

The FWHM values were estimated as 1900, 1200 and 1600~km~s$^{-1}$ for the [Ca II], [Fe II] and Ca~II~NIR lines, respectively. 
The nebular lines have slightly diverse velocity shifts: about $+170$~km~s$^{-1}$ for [Fe~II] $\lambda\lambda$7155, 7453, $-270$~km~s$^{-1}$ for [Ca~II] $\lambda\lambda$7291, 7324 and $-180$~km~s$^{-1}$ for [Ni~II] $\lambda$7376 (the lastter is difficult to measure as it is in the wing of [Ca~II]). As already noted by \cite{foley:2013} for SNe 2002cx, 2005hk, 2008A and 2008ge, the [Fe~II] and [Ca~II] features have shifts in opposite directions, highlighting a quite complex velocity structure of SNe~Iax. 


\subsection{Near-infrared spectral sequence}\label{nir}

\begin{figure*}
\includegraphics[scale=.6,angle=0]{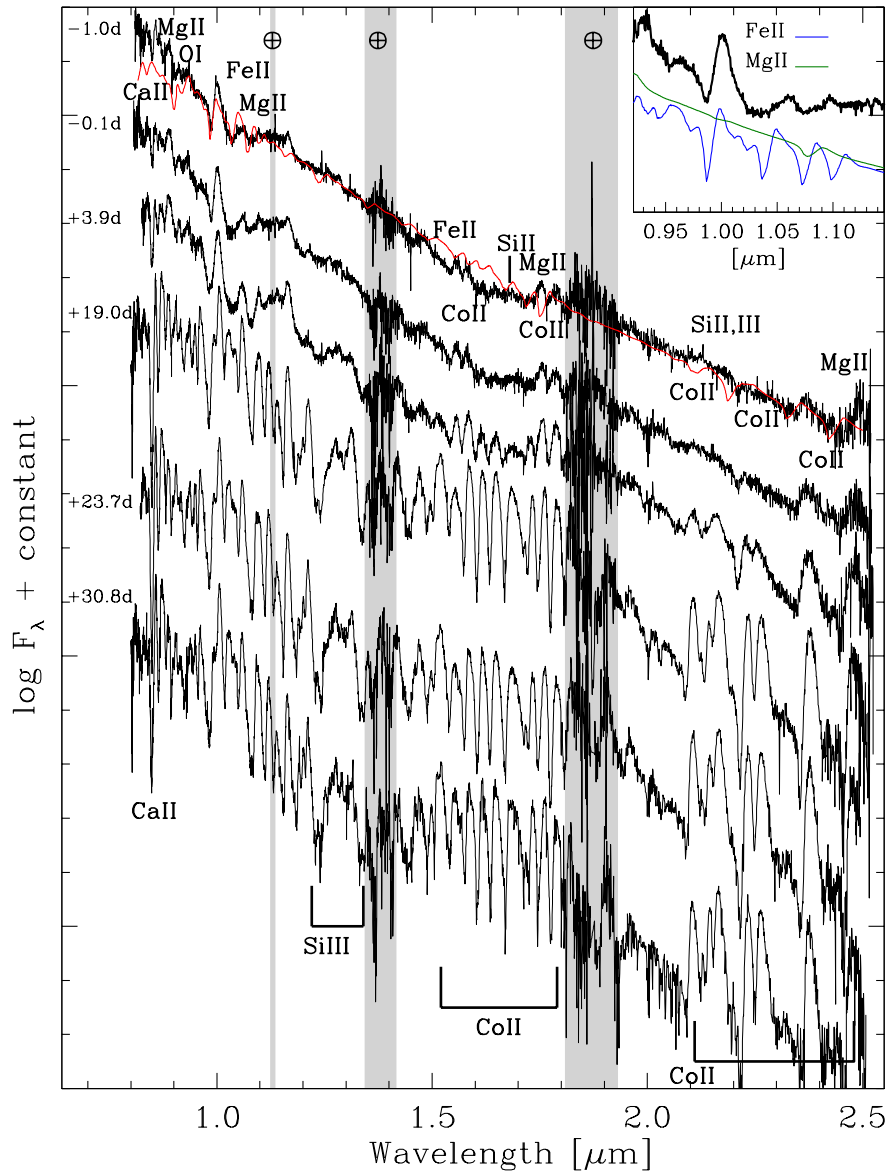}
\caption{NIR spectra of SN~2014ck obtained with the Gemini North Telescope (+ GNIRS). The phase relative to $V$-band maximum is labeled for each spectrum. Prevalent features attributed to Fe~II, Mg~II, Ca~II, Co~II and Si~III are indicated with labels. Telluric regions are indicated with the $\oplus$ symbol and vertical grey bands. 
The $-1$~d spectrum is compared to our best-fit {\tt SYNOW} synthetic spectrum (red).  
The inset on the top shows the $+0.4$~d spectrum (black) in the range 0.95 to 1.15~\micron, showing the main features due to Fe~II (blue) and Mg~II (green). (A colour version of this figure is available in the online journal).} 
\label{spec_nir}
\end{figure*}

\begin{figure}
\includegraphics[scale=.43,angle=0]{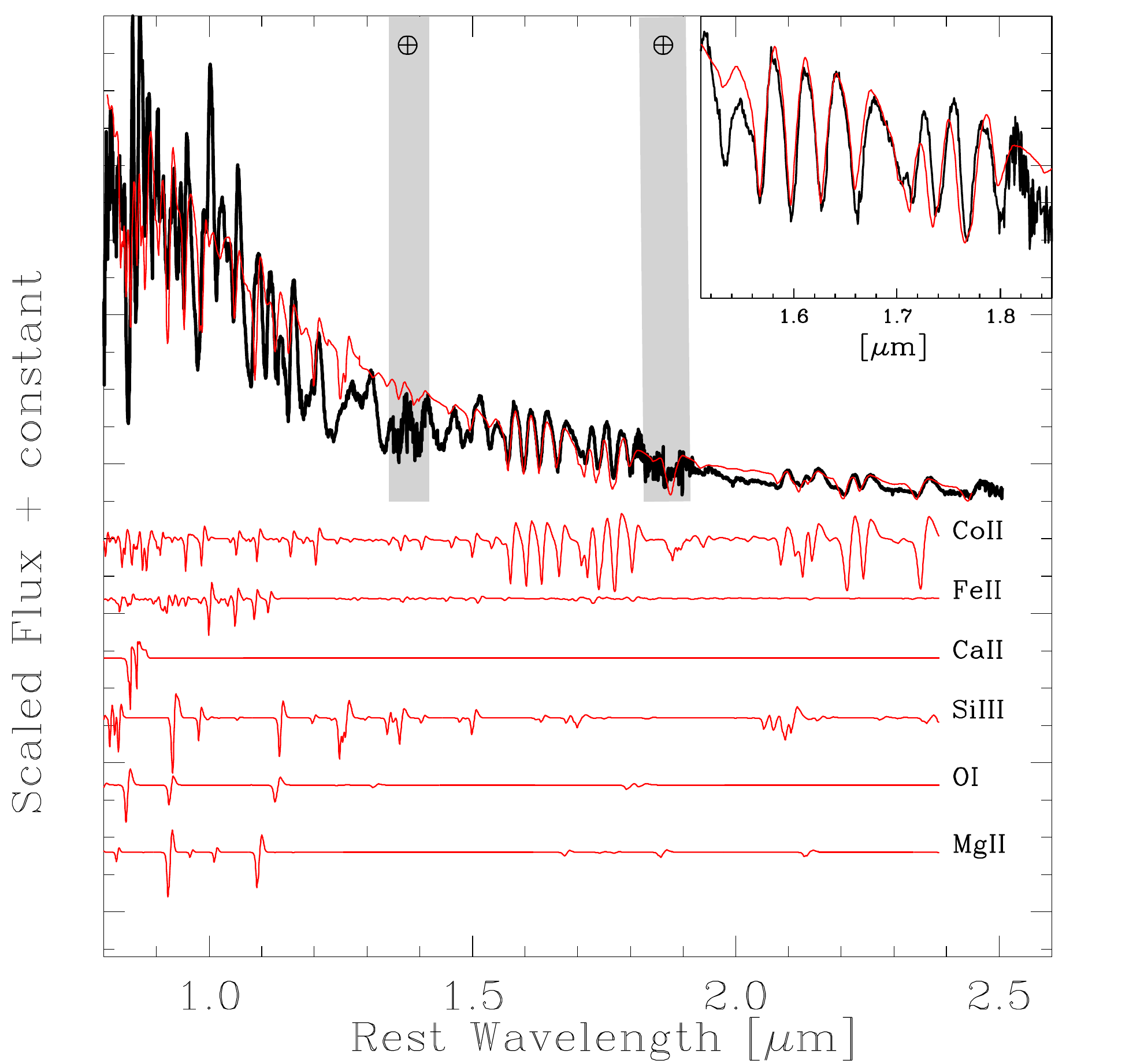}
\caption{NIR spectrum of SN~2014ck at $+19$~d (black) and our best-fit {\tt SYNOW} synthetic spectrum (red). The contribution of prevalent ions is also shown. Telluric regions are indicated with the $\oplus$ symbol and vertical grey bands. Inset: close-up of the $H$-band spectral region, showing the ubiquitous signature of Co~II. (A colour version of this figure is available in the online journal). } 
\label{synow_nir}
\end{figure}

The NIR spectral evolution of SN~2014ck is presented in Figure~\ref{spec_nir}. Before maximum light, large parts of the spectra resemble the infrared tail of a hot blackbody continuum, with the exception of few spectral features around 1~\micron\ and humps between 1.5 and 1.8~\micron.
From $-0.1$~d to $+3.9$~d, the most prominent features are attributed to Fe~II, in particular the stronger line with a  P~Cygni profile at $\sim 1$~\micron\ is Fe~II 0.9998~\micron. 
Mg~II 0.9227 and 1.0927~\micron\ \citep[and possibly weaker lines around 
2.4~\micron\ due to Mg~II transitions, i.e. 2.4041, 2.4044,  2.4125~\micron, see][]{hoeflich:2002} produce shallow notches partially blended with Fe.
Moreover, also around 1~\micron, there might be traces of C~I lines at 0.9093, 0.9406, 1.0693 and 1.1754~\micron, which have been tentatively identified both in the sub-luminous Type~Ia SN~1999by \citep{hoeflich:2002} and in the Type~Iax SN~2012Z \citep{stritz:2015}, in addition to normal SNe~Ia (Hsiao et al. 2013, 2015). However, no confident carbon detections can be made in SN~2014ck NIR spectra. O~I 0.9264~\micron\ should be blended with Mg~II 0.9227~\micron. Actually, in our optical spectra the O~I~$\lambda$7773 -- which is expected to be 3--20 times stronger than the 0.9264~\micron\ line -- is already weak, so the O~I 0.9264~\micron\ is not expected to be a strong feature. 
The 0.8446~\micron\ O~I line may contribute to the absorption, dominated in later phases by the Ca~II NIR triplet. Signatures of Si~II might be present blueward of 1~\micron\ (0.9413~\micron) and the 1.6930~\micron\ Si~II line may be part of the hump at these wavelengths, together with Mg~II (1.6787~\micron) and emerging Co~II lines. 
A {\tt SYNOW} fit (adopting $T_{\rm bb} = 5800$~K and $v_{\rm ph} = 2500$~km~s$^{-1}$, see Section~\ref{bb}) of the $-1$~d spectrum was used to assist for the above line identification, including an extended set of ions (C~I, C~II, O~I, Mg~I, Mg~II, Si~II, Si~III, Ca~II, Fe~II, Fe~III and Co~II, see Figure~\ref{spec_nir}).  
The inset of Figure~\ref{spec_nir} shows the most prominent feature in the earliest spectrum, attributed to Fe~II~0.9998~\micron.
 
Three weeks later, the NIR spectrum radically changes, being strongly dominated by Co~II lines, as previously documented  in all other SNe~Iax with similar data \citep{kromer:2013,stritz:2014,stritz:2015}. Co~II clearly contributes with numerous lines between 1.6 and 1.8~\micron\ soon after maximum light, as it is already present in the spectrum taken at $+3.9$~d. Spectra obtained at $+19$~d or later show distinct absorption at the location of several Co~II lines, most prominently at 1.5759, 1.6064, 1.6361, 1.7772, 1.7462, 2.2205, 2.4596, 2.3613~\micron.
The increasing strength of Co~II with time is attributed both to a lower opacity and a higher abundance of $^{56}$Co in the external ejecta compared with SNe~Ia \citep{hsiao:2013}.
The {\tt SYNOW} fit to the $+19$~d spectrum of SN~2014ck is plotted in Figure~\ref{synow_nir}. We adopted $T_{\rm bb} = 4000$~K and  $v_{\rm ph} = 1900$~km~s$^{-1}$ (see Section~\ref{bb}) and included a smaller subset of the above IMEs and Fe-group ions (Co~II, Fe~II, Ca~II, Si~III, O~I and Mg~II). While it is confirmed that Co~II largely dominates the spectrum redward of 1.5~\micron\ ($H$- and $K$-bands), Fe~II prevails blueward of that wavelength, as in the spectra of SNe~Ia during the transition from the photospheric to the nebular phase \citep{friesen:2014}. 
The Ca~II~NIR triplet is also a prevalent feature starting from the spectrum at $+19$~d, and 
Si~III might help to improve the fit of the features between 1.1 and 1.4~\micron\ \citep[as in SN~2012Z, see][]{stritz:2015}.


The NIR spectral evolution of SN~2014ck is reminiscent of what is observed in SN~2010ae \citep[see][their Figure~4]{stritz:2014} and, overall, in normal SNe~Ia \citep{marion:2009,hsiao:2013,hsiao:2015}. For both SNe~Ia and SNe~Iax the characteristic $H$- and $K$-band iron-peak complex rapidly emerges soon after $B$-band maximum and becomes the dominant feature in the NIR.

\subsection{A summary of the overall spectral characteristics}

Similarly to SN~2008ha \citep{foley:2010b}, the pre-maximum spectra of SN~2014ck show the signatures of Si~II $\lambda\lambda$6347, 6371, S~II $\lambda\lambda$5454, 5640 (together with Ca, O, Na and Fe) and no sign of hydrogen or helium. 
As in SN~2008ha, unburned (C+O) material, specifically C~II $\lambda\lambda$6580, 7234 is also detected, 
as is C~III $\lambda$4647. 
Such analogies in the early phases might suggest they share analogous physical conditions and composition. 
However, unlike SN~2008ha, the late spectrum of SN~2014ck is dominated by iron-peak elements, with the presence of both permitted Fe~II and forbidden [Fe~II], [Fe~III] and [Co~II] emission lines. 
It also displays relatively strong features of both the Ca~II NIR triplet and [Ca~II] $\lambda\lambda$7291, 7324, which have a comparable strength in SN~2008ha but are weak or absent in SN~2002cx. In the NIR spectral region, 
SN~2014ck shows features typical of SNe~Iax, with the ubiquitous presence of Co~II. 
Fe~II lines are present in SN~2014ck at higher expansion velocities than IMEs and unburned C+O elements. Before maximum, Ca~II lines velocities are bracketed by measures of Fe~II, Fe~III and S~II.

\section{Concluding remarks}\label{discussion}

\begin{figure}
\includegraphics[scale=.58,angle=0]{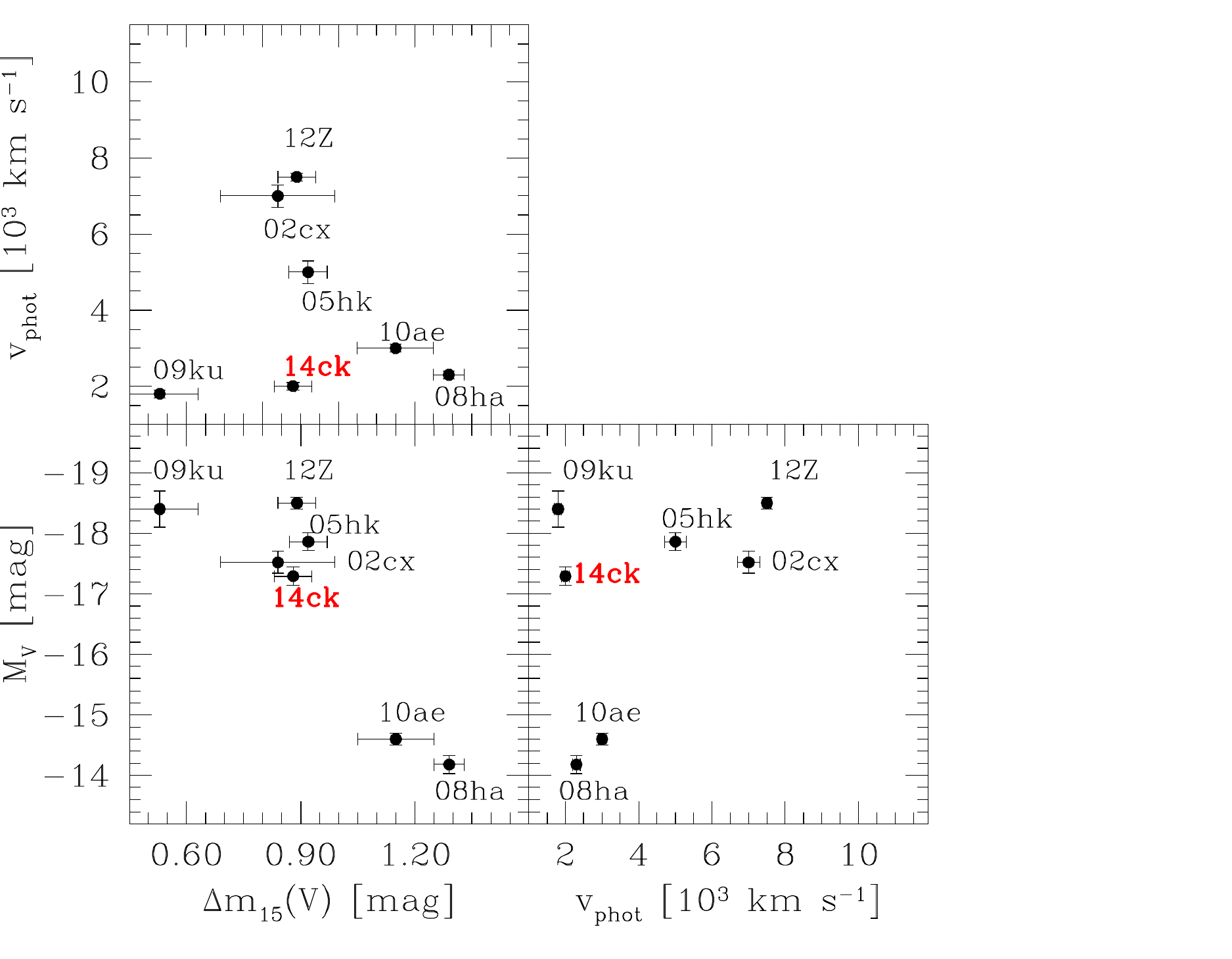}
\caption{Scatter plots of $\Delta m_{15} (V)$, $M_V$ and photospheric velocities $v_{\rm ph}$ derived around +10~days for several well studied SNe~Iax (see Narayan et al. 2011, their Figure~3).} 
\label{conf}
\end{figure}

Mirroring the behaviour already observed for SN~2009ku by \cite{narayan:2011}, the high luminosity and low ejecta velocity for SN~2014ck is contrary to the trend generally seen in the heterogeneous group of peculiar SNe~Iax. In fact, from a photometric point of view SN~2014ck resembles SN~2002cx, the prototype of the class, showing similar peak luminosities ($M_B=-17.37 \pm 0.15$~mag for SN~2014ck vs.\ $M_B=-17.53 \pm 0.26$~mag for SN~2002cx) and decline rates ($\Delta m_{15}(B) = 1.76 \pm 0.15$~mag vs.\ $\Delta m_{15}(B) = 1.7 \pm 0.1$~mag) in all bands. 
Given the almost identical bolometric luminosities at maximum, the synthesised mass of $^{56}$Ni is nearly the same for both objects, $\sim 0.1 M_\odot$.  

Yet despite the relatively high peak luminosity and slow light curve decline, the spectra of SN~2014ck are more similar to those of the three-magnitudes-fainter SNe~2008ha \citep{foley:2009,valenti:2009}, 2007qd \citep{mcclelland:2010} and  2010ae \citep{stritz:2014}. 
All these objects exhibit narrow spectral lines indicating low expansion velocities of the ejecta ($v_{\rm ph}$ from 2500 to 3000~km~s$^{-1}$ at maximum from the absorption minima of Si~II). 
\cite{mcclelland:2010} and \cite{foley:2013} suggested the existence of a relation between peak luminosity, ejecta velocity and light-curve shape, with higher-velocity SNe~Iax being more luminous and more slowly declining. 
On this basis, they argued that the full class of SNe~Iax might originate from a single explosion mechanism. 

Both SNe~2009ku and 2014ck, with their low velocity and relatively high luminosity, are outliers in these relations, and the results of this work confirm and reinforce the results of \cite{narayan:2011}.  
This is illustrated in Figure~\ref{conf}, which shows the positions of a sample of SN~Iax in a three-variable phase space composed of $\Delta m_{15}(V)$, $M_V$ and $v_{\rm ph}$ measured 10 days after maximum \cite[see][their Figure~3]{narayan:2011}.
In fact, this sample of SNe~Iax shows: 
(i) there is not a monotonic correlation between $v_{\rm ph}$ and $\Delta {\rm m}_{15}(V)$, as at low velocities the spread in the decline rate is wide (top-left panel); 
(ii) there is evidence of a linear correlation between $\Delta {\rm m}_{15}(V)$ and $M_V$  (bottom-left); and 
(iii) SN~2014ck (and also SN~2009ku) are at odd with the suggestion of \cite{mcclelland:2010} that higher velocity SNe~Iax also have higher peak luminosities (bottom-right panel). 
We note that these are still small-number statistics, and therefore these findings need further confirmation. However, at present we cannot conclude SNe~Iax are a one-dimensional sequence from fainter to brighter events. 

Comprehensive explosion models are required to describe both the diversity and homogeneity among SNe~Iax. On the whole, the observations seem to indicate that they may originate from a homogeneous population \citep{foley:2013}. 
However, specifically, there appears to be a diverse range of rise times, peak luminosities, expansion velocities, etc., and, as underlined above, there is not a clear correlation among these physical parameters. Thus, there may be multiple progenitor paths and explosion mechanisms to create a SN~Iax. 
To date, the progenitors of SNe~Iax are still a subject of debate \citep{valenti:2009,foley:2009,foley:2010b,moriya:2010,liua:2015}. The search and detection of progenitor stars or systems in pre-explosion images is, in principle, a promising technique to test different progenitor models \citep{mccully:2014,liua:2015,liub:2015}. For SN~2014ck, archival {\it HST} images were obtained and no progenitor was detected at the SN position. The available images are not very deep and provide a $3\sigma$ limit of $M_{\rm F625W} > -6.5$~mag. This limit allows us to rule out only the most-luminous Wolf-Rayet stars as a potential progenitor.

Overall, the optical and NIR observational features summarised above favour a thermonuclear explosion of a C/O WD for SN~2014ck. A failed deflagration model \citep{jordan:2012,kromer:2013} might explain most of its observed properties, from the low peak luminosity and energetics (see Section~\ref{bolometric}) to the spectral characteristics (see Section~\ref{spec_evol}). In particular, the moderate $M_{\rm Ni}$ production derived for this event and the low expansion velocities are well matched in this scenario. The failed deflagration is too weak to unbind the WD, leaving behind a gravitationally bound, compact remnant around $\sim 1 M_\odot$. The low ejecta mass of SN~2014ck is partially consistent with that prediction. However, the rise time predicted by these models \citep[see Table~4 in][]{fink:2014} is too fast compared to the rise time derived for SN~2014ck ($t_{\rm rise} = 16.9^{+4.5}_{-2.7}$~days, see Section~6.2). Above all, the modelled burning of a C/O WD via a turbulent deflagration flame produces a homogeneous mixing of elements in the ejecta, with unburned material, partially burned material and fully burned (to the iron peak) material throughout the ejecta. Significant mixing in the ejecta was suggested for SN~2005hk by \cite{phillips:2007}. The failed deflagration model was also considered the most probable scenario for SN~2012Z by \cite{yamanaka:2015}. On the contrary, for the same event, \cite{stritz:2015} pointed out evidence of a layered structure for calcium, silicon and magnesium, which is instead the signature of a detonation. Thus, they suggested that these elements are produced in a detonation phase after the mixing has already occurred and the majority of the iron peak elements have been produced in a previous deflagration phase, i.e. a pulsational delayed detonation \citep[PDD;][]{hoeflich:1995,hoeflich:1996}.   

Looking at the velocity distribution of elements in SN~2014ck (Figure~12), 
the presence of Fe~II features up to high velocities, as reported by \cite{stritz:2015} for SN~2012Z (and also for few normal SNe~Ia), seems to suggest 
a layered structure in the ejecta, arguing in favour of a detonation phase, which might have followed a deflagration. On the other hand, C+O elements and Si~II are moderately mixed, and one could argue that they would be identified at higher velocities if earlier spectra were available. Moreover, explaining iron-group material in the outer layers is still a challenge for explosion models, even if some results have been obtained within a delayed-detonation scenario \citep[see for example][and references therein]{hach:2013}. So far, several questions are still open, among them the mechanism of transition from deflagration to detonation. To be thorough, it is difficult to explain within PDD models the extremely low photospheric velocity at maximum of SN~2014ck. We also underline, once again, that the severe blending across the spectra of SNe~Iax, might prevent secure line identifications, as was detailed by \cite{szalai:2015}. Consequently, the derived expansion velocities of the elements is ill constrained, as well as the distinction between a layered or a mixed structure of the ejecta. 

The analysis of our extended data set of SN~2014ck cannot exclude either the failed deflagration or the PDD models for this event. In addition, the comparison with other SNe~Iax highlights that the diversity within this class of transients cannot be reduced to a one-parameter description. This may also imply that distinct progenitors and/or explosion mechanisms are involved, despite the overall similarity of the main observables.

\section*{Acknowledgments}

This paper is based on observations made with: the LCOGT 2.0~m (Haleakala, Hawaii, USA) and 1.0~m (McDonald Observatory, Texas, USA) telescopes; 
the 2.56~m Nordic Optical Telescope (La Palma, Spain); the INAF Osservatorio Astronomico di Padova Copernico 1.82~m Telescope (Mt.~Ekar, Asiago, Italy); the 3.58~m Telescopio Nazionale Galileo 
(La Palma, Spain); the 8.1~m Gemini-N Telescope (Hilo, Hawaii, USA); and the 10.4~m Gran Telescopio Canarias (La Palma, Spain).
We acknowledge the staff at INAF OAPd in Asiago and at LCOGT for their support.

Based on observations obtained at the Gemini Observatory under Program
ID GN-2014A-Q-8 and GN-2014B-Q-41. Gemini is operated by the
Association of Universities for Research in Astronomy, Inc., under a
cooperative agreement with the NSF on behalf of the Gemini partnership: the National Science
Foundation (United States), the National Research Council (Canada), CONICYT
(Chile), the Australian Research Council (Australia), Minist\'{e}rio da Ci\^{e}ncia,
Tecnologia e Inova\c{c}\~{a}o
(Brazil) and Ministerio de Ciencia, Tecnolog\'{i}a e Innovaci\'{o}n
Productiva (Argentina).
Also based on observations made with the NASA/ESA Hubble Space Telescope, obtained from the data archive at the Space Telescope Science Institute. STScI is operated by the Association of Universities for Research in Astronomy, Inc. under NASA contract NAS 5-26555.

We thank WeiKang Zheng and Alex Filippenko for sending us pre-discovery and discovery Lick/KAIT images of UGC~12182 from which non-detection upper limits of SN~2014ck were determined and the rise time estimated. We also thank the anonymous referee for the thorough review of the paper.

AP, SB, NER, LTar, GT and MT are partially supported by the PRIN-INAF 2014 with the project ``Transient Universe: unveiling new types of stellar explosions with PESSTO". 
NER acknowledge the support from the European Union Seventh Framework Programme (FP7/2007-2013) under grant agreement n.\ 267251 ``Astronomy Fellowships in Italy'' (AstroFIt). AMG acknowledges financial support by the Spanish {\it Ministerio de 
Econom\'ia y Competitividad} (MINECO), grant ESP2013-41268-R. 
MS and EYH acknowledge support provided by the Danish Agency for Science and Technology and Innovation through a Sapere Aude Level 2 grant.

This research has made use of: the NASA/IPAC Extragalactic Database (NED) which is operated by the Jet Propulsion Laboratory, California Institute of Technology, under contract with the National Aeronautics and Space Administration;
 {\sc iraf} packages, distributed by the National Optical Astronomy Observatory, operated by the Associated Universities for Research in Astronomy, Inc., under cooperative agreement with the National Science Fundation.

\label{lastpage}
\end{document}